\documentclass[preprint,12pt,authoryear]{elsarticle}
\usepackage{lineno}
\usepackage{amsmath}
\usepackage{ae,aecompl}
\usepackage[]{graphicx}
\usepackage{amsfonts}
\usepackage{amssymb}
\usepackage{epstopdf}
\usepackage{epsfig}
\usepackage{float}
\def\lsim{\lower.5ex\hbox{$\; \buildrel < \over \sim \;$}}
\def\gsim{\lower.5ex\hbox{$\; \buildrel > \over \sim \;$}}
\renewcommand{\d}{\mathrm{d}}

\journal{New Astronomy}

\begin{document}

\begin{frontmatter}

\title{Influence of matter geometry on shocked flows-I: Accretion in the Schwarzschild metric}

\author[Tarafdar]{Pratik Tarafdar}
\address[Tarafdar]{S. N. Bose National Centre For Basic Sciences, Block-JD, Sector III, Salt Lake, Kolkata-700106, India.}
\ead[Tarafdar]{pratik.tarafdar@bose.res.in}
\author[Das]{Tapas K. Das}
\address[Das]{Harish-Chandra Research Institute, Chhatnag Road, Jhunsi, Allahabad, India.}
\ead[Das]{tapas@hri.res.in}


\begin{abstract}
\noindent
This work presents a comprehensive and extensive study to illustrate how the geometrical configurations of low angular 
momentum axially symmetric general relativistic matter flow in Schwarzschild metric may influence the formation of 
energy preserving shocks for adiabatic/polytropic accretion as well  as of temperature preserving dissipative shocks for 
for the isothermal accretion onto non-rotating astrophysical black holes. The dynamical and thermodynamic states of post 
shock polytropic and isothermal flow have been studied extensively for three possible matter geometries, and it has been 
thoroughly discussed about how such states depend on the flow structure, even when the self gravity and the back 
reaction on the metric are not taken into account. Main purpose of this paper is thus to mathematically demonstrate 
that for non-self gravitating accretion, various matter geometry, in addition to the corresponding space-time 
geometry, controls the shock induced phenomena as observed within the black hole accretion discs. This work is expected 
to reveal how the shock generated phenomena (emergence of the outflows/flare or the behaviour of QPO of the associated 
light curves) observed at the close proximity of the horizon depends on the physical environment of the source 
harbouring a supermassive black hole.
\end{abstract}


\end{frontmatter}

\section{Introduction}
Low angular momentum accretion flow onto astrophysical black holes exhibit multi-transonic features 
(\cite{lt80apj,az81apj,mp82aa,fukue83pasj,lu85aa,lu86grg,mc86aa,blaes87mnras,chakrabarti89apj,ak89apj,
das02apj,fukue04pasj,fukue04pasja,mdc06mnras,dc12na}). 
That stationary shock may form for such flow 
configuration, was first ever discussed by Fukue (\cite{fukue87pasj}) for general relativistic flow in the Kerr metric. Following 
Fukue's prescription, several authors made contribution to study the shocked block hole accretion for Newtonian, post-Newtonian 
and general relativistic flow onto rotating and non-rotating astrophysical black holes 
(\cite{chakrabarti89apj,nf89pasj,ac90apj,trft92apj,ky94mnras,sm94mnras,yk95aa,pariev96mnras,nayakama96mnras,pa97mnras,btb97prl,
tkb98aa,das02apj,dpm03apj,ft04apj,ottm04pasj,tgfrt06apj,ny08apj,ny09apj,dc12na}).
Aforementioned works have been performed to study the shocked flow in individual flow geometries and no comprehensive study 
has been carried out to understand the role of geometric configuration of accretion flow in influencing the shock 
related phenomena. \\

Non spherical accretion process onto astrophysical black holes are usually studied for three different geometric configurations 
of axially symmetric flow. In simplest possible set up, the thickness of the disc is assumed to be space invariant 
(constant for all radial distance) for stationary state. All planes are thus considered to be equatorial plane since a symmetry about 
two axes exists in such model. The flow structure can thus be mapped with a right circular cone with constant half thickness. In 
next variant, the flow is considered to be quasi-spherical in shape where the ratio of the radial distance at a point and the local 
half thickness at that particular point remains invariant for all $r$. All directions are equivalent to the radial direction for this model. 
Quasi-spherical flow, or the conical flow, as is it mentioned in the literature 
(see, e.g. section 4.1 of \cite{cd01mnras,nard12na,bcdn14cqg}), is considered 
to be idealmost to model low angular momentum inviscid accretion since weakly rotating advection dominated accretion is best described by such 
geometric structure. A rather non-trivial axially symmetric accretion configuration requires the matter to be in hydrostatic equilibrium 
along the vertical direction. The local flow thickness becomes non linear function of radial distance, as well as of the local radial sound speed. 
We believe that it is instructive to investigate the properties of the shocked accretion flow for aforementioned flow configurations in a unified 
manner so that the influence of the geometric configuration of matter on the general relativistic accretion dynamics can be understood. This is 
precisely the main objective of our work. In the present work, which is the first paper of our series, we study the shock formation phenomenona 
for general relativistic accretion in the Schwarzschild metric for three different flow structures, and will investigate how the flow structure 
affects the properties of the post shock flow. In our next work, we intend to continue such approach for flows in the Kerr space-time in order to 
incorporate the role of the black hole spin in governing the shocked accretion flow for all three different flow configurations. 
The corresponding expressions for the flow thickness in all the configurations in the existing literature, however, are derived from 
a set of idealised assumptions. In reality, the rigorous derivation of the flow thickness may be accomplished by using the framework 
of non-LTE radiative transfer (e.g. \cite{hh98apj,dh06apjs}) or, by using the Grad-Shafranov equations for the MHD-related 
aspects of the flow (e.g., \cite{beskin97pu,bt05aa,beskin09}). For all three flow geometries, we will consider the flow 
along the equatorial plane only. Low angular momentum inviscid flow will be assumed to have considerable radial advective velocity to begin with. 
For low angular momentum advective accretion flow, large radial velocity close to the black holes implies $\tau_{inf}<<\tau_{visc}$, where 
$\tau_{inf}$ and $\tau_{visc}$ are the infall and the viscous time scales, respectively. Large values for radial components of 
the velocities even at large distances are due to low angular momentum content of the flow (\cite{bi91mnras,ib97mnras,pb03apj}). 
The concept of low angular momentum flow (capable of providing the favourable configuration of the formation of 
standing shock) is not a theoretical abstraction. Sub-Keplerian flows are observed in 
accreting detached binaries fed by OB stellar winds (\cite{is75aa,ln84ssr}), semi-detached low-mass 
non-magnetic binaries (\cite{bbck98mnras}), and super-massive black holes accreting from stellar clusters rotating with relatively 
low angular velocities near the galactic centre (\cite{illarionov88sa,ho99} and reference therein). 
Moreover, even turbulence may lead to low angular momentum flow for a standard Keplerian disc (see \cite{ia99mnras} e.g., and references therein). 
In what follows, we will derive the fluid dynamic equations (general 
relativistic Euler and continuity equations) from the stress energy tensor of an ideal fluid, and will argue how such equations govern the 
accretion flow. We will then provide the stationary integral transonic solutions of such equations and discuss how one obtains the multi-transonic 
flow profile by incorporating an energy preserving Rankine Hugoniot type standing shock for the adiabatic flow and a temperature preserving standing 
shock for isothermal flow, for three different flow geometries. The properties and structure of the post shock flow will be described to demonstrate 
how the matter geometry determines the flow profile in general.

\section{Configuration of the background fluid flow}
\noindent

A (3+1) stationary axisymmetric space-time is considered with two
commuting Killing fields. The local timelike Killing field $\xi^{\mu}\equiv\left(\partial{/\partial{t}}\right)^\mu$ 
generates stationarity and $\phi^{\mu}\equiv\left(\partial{/\partial{\phi}}\right)^\mu$ generates axial symmetry. 
All distances and velocities are scaled in units of $GM_{BH}/c^2$ and $c$ respectively, where $M_{BH}$ is the black hole mass. 

To describe the flow structure, the energy momentum tensor 
of an ideal fluid of the form
\begin{equation}
T^{\mu\nu}= (\epsilon+p)u^{\mu}u^{\nu}+pg^{\mu\nu}
\label{Tmunu}
\end{equation}
is considered in a Boyer-Lindquist (\cite{bl67jmp}) line element normalized for $G=c=M_{BH}=1$ and 
$\theta=\pi/2$ (equatorial plane) as defined below (\cite{nt73})
\begin{equation}
ds^2=g_{{\mu}{\nu}}dx^{\mu}dx^{\nu}=-\frac{r^2{\Delta}}{A}dt^2
+\frac{A}{r^2}\left(d\phi-\omega{dt}\right)^2
+\frac{r^2}{\Delta}dr^2+dz^2\,
\label{ds2}
\end{equation}
where
\begin{equation}
\Delta=r^2-2r+a^2, A=r^4+r^2a^2+2ra^2,\omega=2ar/A\,
\label{Delta}
\end{equation}
$a$ being the Kerr parameter.
Hence, for a non-rotating black hole where $a=0$ the required metric elements are:
\begin{equation}
g_{rr}=\frac{r^2}{\Delta},~g_{tt}=-\frac{r^2{\Delta}}{A},~
g_{\phi\phi}=\frac{A}{r^2},~ g_{t\phi}=g_{\phi{t}}=0,.
\label{gmunu}
\end{equation}
The specific angular
momentum $\lambda$ (angular momentum per unit mass)
and the angular velocity $\Omega$ can thus be expressed as
\begin{equation}
\lambda=-\frac{u_\phi}{u_t}, \;\;\;\;\;
\Omega=\frac{u^\phi}{u^t}
=-\frac{g_{t\phi}+\lambda{g}_{tt}}{{g_{\phi{\phi}}+\lambda{g}_{t{\phi}}}}\, .
\label{Omega1}
\end{equation}

The equation of state, 
\begin{equation}
p=K\rho^{\gamma} 
\label{eqnofstatepoly}
\end{equation}
describes polytropic accretion, where the polytropic index 
$\gamma$ is assumed to be constant for the steady state flow. 
The calculations have been performed over a wide range of $\gamma$.
Thus, all polytropic indices of astrophysical relevance have been covered.
The proportionality constant $K$ in eq. (\ref{eqnofstatepoly}) measures 
the specific entropy of the accreting fluid when additional entropy is not generated. 
The specific enthalpy $h$ is formulated as
\begin{equation}
h=\frac{p+\epsilon}{\rho},
\label{enthalpy1}
\end{equation}
where $\epsilon$ is the energy density (including rest mass density) 
and the internal energy is given by
\begin{equation}
\epsilon=\rho +\frac{p}{\gamma-1}
\label{epsilon}
\end{equation}
The adiabatic sound speed $c_s$ is defined by
\begin{equation}
c_s^2=\left(\frac{\partial p}{\partial \epsilon}\right)_{\rm constant~enthalpy}
\label{csq1}
\end{equation}
The enthalpy at constant entropy is given by
\begin{equation}
h = \frac{\gamma - 1}{\gamma - \left(1 +c_{s}^2\right)}
\label{enthalpy3}
\end{equation}

We have also investigated the isothermal accretion flow, where the equation of state is given by 
\begin{equation}
p=\rho{c_s^2}=\frac{\cal R}{\mu}\rho{T}=\frac{\rho{\kappa_B}T}{{\mu}m_H}
\label{eqnofstateiso}
\end{equation}
${\cal R}$ is the universal gas constant, $K$ is the specific entropy, $\kappa_B$ is the Boltzmann constant, 
$T$ is the flow temperature, $\mu$ is the mean molecular mass of fully ionized hydrogen, $m_H$ is Hydrogen atom mass and $c_s$ is the 
position independent sound speed, respectively.

\section{The first integrals of motion}
\noindent
The general relativistic Euler equations for polytropic flow are obtained from
\begin{equation}
T^{\mu\nu}_{;\nu}=0.
\label{Euler1}
\end{equation}
and the continuity equation is given by
\begin{equation}
\left(\rho u^{\mu}\right)=0.
\label{continuity}
\end{equation}

\subsection{Integral solution of the linear momentum conservation equation}
\noindent
Contracting eq. (\ref{Euler1}) with $\phi^{\mu}$ one obtains 
the angular momentum per baryon, $hu_{\phi}$, which is conserved. 
Contraction of eq. (\ref{Euler1}) with $\xi^{\mu}$ gives $hu_t$, which is the relativistic 
Bernoulli's constant and is conserved (\cite{anderson89mnras}).
It can be identified with ${\cal E}$, the total specific energy of the ideal GR 
fluid (see, e.g., \cite{dc12na} and references therein) scaled in units of the rest mass energy.

The specific angular momentum is defined as 
\begin{equation}
\lambda = -\frac{u_{\phi}}{u_{t}}
\label{lambda1}
\end{equation}
The angular velocity $\Omega$ is expressed in terms of $\lambda$ as
\begin{equation}
\Omega=\frac{u^{\phi}}{u^t}=-\frac{g_{t\phi}+g_{tt}\lambda}{g_{\phi\phi}+g_{t\phi}\lambda}
\label{Omega2}
\end{equation}
From the normalization condition $u^{\mu}u_{\mu}=-1$ one obtains 
\begin{equation}
u_t=\sqrt{\frac{g_{t\phi}^2-g_{tt}g_{\phi\phi}}{(1-\lambda\Omega)(1-u^2)(g_{\phi\phi}+\lambda g_{t\phi})}}
\label{ut1}
\end{equation}

The corresponding expression for conserved energy ${\cal E}$ is therefore given by
\begin{equation}
{\cal E} = \frac{\gamma - 1}{\gamma - \left(1 +c_{s}^2\right)}
           \sqrt{\frac{g_{t\phi}^2-g_{tt}g_{\phi\phi}}{(1-\lambda\Omega)(1-u^2)(g_{\phi\phi}+\lambda g_{t\phi})}}
\label{E2}
\end{equation}
It is clear that the expression for ${\cal E}$ depends on space-time geometry. It does not depend on 
matter geometry. Since the flow has been assumed to be non self-gravitating, hence the accreting fluid does not backreact 
on the space-time metric itself. In case of isothermal flow, energy gets dissipated in order to maintain a constant 
temperature. Thus, the total specific energy is not 
conserved. Integration of the relativistic Euler equation leads to an entirely different algebraic form 
for the first integral of motion, which cannot be identified with the total energy of the system.
 
The isotropic pressure, which is proportional to the energy density, is given by
\begin{equation}
p = c_s^2 {\epsilon}
\label{eqnofstateiso1}
\end{equation}
(see, e.g., \cite{ydl96ap} and references therein)
From the time part of eq. (\ref{Euler1}), we obtain 
\begin{equation}
\frac{\d u_t}{u_t}=-\frac{\d p}{p+\epsilon}
\label{anal32}
\end{equation}
\begin{equation}
\frac{\d u_t}{u_t}=-\frac{1}{h}\frac{\d p}{\d \rho}\frac{\d \rho}{\rho}
\label{anal33}
\end{equation}
Since,  
\begin{equation}
c_s=\sqrt{\frac{1}{h}\frac{\d p}{\d \rho}}
\label{anal34}
\end{equation}
one obtains 
\begin{equation}
\ln u_t=-c_s^2\ln \rho + A, \textrm{ where $A$ is a constant}
\label{anal35}
\end{equation} 
\begin{equation} 
{u_t}{\rho^{c_s^2}} = {\rm C_{\rm iso}}
\label{anal36}
\end{equation}
${\rm C_{\rm iso}}$ is the first integral of motion in this case. 
It must not be confused with the total conserved specific energy ${\cal E}$. 

\subsection{Integral solution of the mass conservation equation}
\noindent
The mass conservation equation (\ref{continuity}) gives
\begin{equation}
\frac{1}{\sqrt{-g}}(\sqrt{-g}\rho u^{\mu})_{,\mu}=0,
\label{anal37}
\end{equation}
where $g\equiv \det(g_{\mu\nu})$. This implies 
\begin{equation}
\left[(\sqrt{-g}\rho u^\mu)_{,\mu}d^4 x=0\right]
\label{anal38}
\end{equation}
where $\sqrt{-g}d^4 x$ is the covariant volume element. We assume 
that $u^{\theta}$ (in spherical polar co-ordinates)/
$u^z$ (in cylindrical co-ordinates) are relatively negligible when compared 
to the transformed radial component $u^r$. Using this assumption we get
\begin{equation}
\partial_r(\sqrt{-g}\rho u^r)\d r \d\theta\d\phi = 0,
\label{anal39}
\end{equation}
for stationary flow in spherical polar co-ordinates and 
\begin{equation}
\partial_r(\sqrt{-g}\rho u^r)\d r \d z\d\phi = 0,
\label{anal40}
\end{equation} 
in cylindrical co-ordinates.

Eq. (\ref{anal39}) is integrated for $\phi$ ranging from $0$ to $2\pi$ and 
$\theta$ ranging from $-H_{\theta}$ to $H_{\theta}$, where $\pm H_{\theta}$ are 
corresponding values of the co-ordinates above and below the equatorial 
plane respectively, for local half thickness $H$, to 
obtain the conserved quantity $\dot{M}$ which represents the mass accretion 
rate when $\theta=\frac{\pi}{2}$ (i.e. on the equatorial plane). \\

The expression for $\dot{M}$ is different for different matter geometry configurations. 
A generalized expression for $\dot{M}$ may be written as
\begin{equation}
\dot{M} = \rho u^r\mathcal{A}(r)
\label{anal41}
\end{equation}
where $\mathcal{A}(r)$ represents the 2D surface area through which 
the steady state inbound mass flux is calculated.

\section{Stationary transonic solutions for axisymmetric background flow}

\subsection{polytropic accretion}
The explicit expression of specific energy for polytropic accretion in the Schwarzschild metric, 
is of the following form
\begin{equation}
{\cal E} = -\frac{\gamma -1}
{(\gamma -(1+c_s{}^2))}\sqrt{\frac{(1-\frac{2}{r})}{(1-\frac{\lambda ^2}{r^2}(1-\frac{2}{r}))(1-u^2)}}
\label{anal42}
\end{equation}

\subsubsection{Flow with constant thickness}
\noindent
For an accretion disc with a constant height $H$, the mass accretion rate is given by 
\begin{equation}
\dot{M}=2 \pi \rho \frac{u\sqrt{1-\frac{2}{r}}}{\sqrt{1-u^2}} r H
\label{anal43}
\end{equation}

Equations (\ref{anal42} - \ref{anal43}) contain three unknowns 
$u,c_s$ and $\rho$, which are functions of $r$. Hence, one of the three has to 
be eliminated by expressing it in terms of the other two. Here, we are interested in studying the 
profile for the radial Mach number $M=\frac{u}{c_s}$ to locate the acoustic horizons (i.e. $r$ at which $M$ is unity). 
Therefore, we need to substitute $\rho$ in terms of the other related quantities. 
For this purpose, we make use the following transformation- \\
${\dot \Xi}={\dot M}K^{\frac{1}{\gamma-1}}{\gamma^{\frac{1}{\gamma-1}}}$. \\
Applying the definition $c_s^2=\left(\frac{\partial{p}}{\partial{\epsilon}}\right)_{\rm Constant~Entropy}$
and equation of state for the given flow, this transformation leads to
\begin{equation}
\dot{\Xi}=2{\pi} \frac{u\sqrt{1-\frac{2}{r}}}{\sqrt{1-u^2}} rc_s^{\frac{2}{\gamma -1}}(\frac{\gamma -1}{\gamma - (1+c_s^2)})^\frac{1}{\gamma -1}H
\label{anal44}
\end{equation}
Now, it is known that $\sigma$, which is the entropy per particle can be expresed in terms of $K$ and $\gamma$ by (\cite{landauPK})
$$
\sigma=\frac{1}{\gamma -1}\log K+
\frac{\gamma}{\gamma-1}+{\rm constant}
\label{anal45}
$$
where, chemical composition of the fluid determines the value of the constant.
Using the above expression, one can imply that $K$
measures, in this case, the specific entropy of the accreting matter.
Hence, ${\dot {\Xi}}$ may be interpreted as a measure of the total
inbound entropy flux of the fluid and can thus be defined as 
the stationary entropy accretion rate. \\

This entropy accretion rate was proposed for the first time in
\cite{az81apj,blaes87mnras}. It was used to formulate the transonic solutions
for stationary, non-relativistic, low angular momentum, axisymmetric accretion
in the Paczy\'nski - Wiita (\cite{pw80aa}) pseudo-Newtonian potential onto a Schwarzschild black hole. \\

Differentiating eq. (\ref{anal44}), space gradient of the acoustic velocity
can be related to that of the advective velocity by
\begin{equation}
\frac{dc_s}{dr}= -\frac{\gamma -1}{2}\frac{\left\{\frac{1}{u}+
\frac{u}{1-u^2}\right\}\frac{du}{dr}+\left\{\frac{1}{r}+
\frac{1}{r^2(1-\frac{2}{r})}\right\}}{\frac{1}{c_s}+\frac{c_s}{\gamma -(1+c_s)}}
\label{anal46}
\end{equation}
Differentiating eq. (\ref{anal42}) w.r.t. $r$ we get one more equation connecting 
$dc_s/dr$ with $du/dr$. $dc_s/dr$ obtained from
eq. (\ref{anal46}) is then substituted into the second relation and thus, expression for $\frac{du}{dr}$, i.e. 
space gradient of the advective velocity is found to be
\begin{equation}
\frac{du}{dr} = 
\frac{c_s{}^2\left\{\frac{1}{r}+\frac{1}{r^2(1-\frac{2}{r})}\right\}-f_2(r,\lambda )}{(1-c_s{}^2)\frac{u}{1-u^2}-\frac{c_s{}^2}{u}} 
= \frac{\it N_1}{\it D_1}
\label{anal47}
\end{equation}
where, 
\begin{subequations}
\begin{align}
f_2(r,\lambda) = -\frac{\lambda ^2}{r^3}\left\{\frac{1-\frac{3}{r}}{1-\frac{\lambda ^2}{r^2}(1-\frac{2}{r})}\right\}+\frac{1}{r^2(1-\frac{2}{r})} \label{anal48b} \\
\nonumber \\
\text{Another quantity $f_1(r,\lambda)$ is defined to be used later,} \nonumber \\
f_1(r,\lambda)=\frac{3}{r}+\frac{\lambda ^2}{r^3}\left\{\frac{1-\frac{3}{r}}{1-\frac{\lambda ^2}{r^2}(1-\frac{2}{r})}\right\}
\label{anal48a}\\
\nonumber
\end{align}
\end{subequations}
Eqs. (\ref{anal46} -- \ref{anal47}) are observed to be an autonomous dynamical system 
modelled by a set of non-linear first order differential equations,
whose integral solutions are 
phase portraits on the $M$ vs $r$
plane. Setting the numerator and denominator of
eq. (\ref{anal47}) to be zero simultaneously, we obtain the `regular' critical point conditions as,
\begin{equation}
\left[u=c_s\right]_{r_c}, ~~~ \left[c_s\right]_{r_c}=
\sqrt{\frac{f_2(r_c,\lambda )}{\frac{2}{r_c}+\frac{1}{r_c{}^2(1-\frac{2}{r_c})}}}
\label{anal49}
\end{equation}
Since we are dealing with continuous flow of a transonic fluid in physical space, 
we need to look only for the `smooth' or `regular' critical points, for
which the dynamical velocities and their space derivatives are regular
and do not diverge. Such critical points may either be saddle-type
with a transonic solution passing through it, or 
centre-type, through which physical transonic solutions cannot
pass. Other categories also exist, including `singular' critical points (with continuous velocities 
but diverging derivatives). All these categorizations and regularity conditions for a critical point 
associated with an acoustic horizon have been elaborated in \cite{abd06cqg}.

Equation (\ref{anal49}) does not provide the locations of the critical points. One needs to 
solve eq. (\ref{anal42}) employing the critical point conditions for
a given set of values of the system parameters 
${\cal E}$, $\lambda$ and $\gamma$. Obtaining values of $c_s$ and $u$ from eq. (\ref{anal49}) and 
substituting them in eq. (\ref{anal42}) we obtain a polynomial equation for the critical point location 
$r=r_c$. Coefficients of the polynomial equation in $r_c$ are complicated functions of 
${\cal E}$, $\lambda$, $\gamma$, $u_c$ and $c_{sc}$. Thus for a given combination of $\left[{\cal E},\lambda,\gamma\right]$ 
one may numerically solve the polynomial equation for values of $r_c$.
The domains of astrophysically relevant values for ${\cal E}$, $\lambda$ and $\gamma$ are given by 
\begin{equation}
\left[1{\lsim}{\cal E}{\lsim}2,0<\lambda{\le}2,4/3{\le}\gamma{\le}5/3\right]
\label{anal52}
\end{equation}
From the critical point conditions obtained for constant height flow, it is evident that the 
$r_c$ (critical point) $=r_h$ (the acoustic horizon). The nature of such critical points can also be determined using 
the numerical values of $r_c,{\cal E},\lambda$ and $\gamma$ (\cite{gkrd07mnras}). \\

The polynomial equation in $r_c$ can have either no real solution or, at most, three real solutions for a given 
set of $\left[{\cal E},\lambda,\gamma\right]$. The present work does not interest us in the non-transonic flows. 
A single real solution of saddle type implies monotransonic flow. Two saddle type solutions indicate a homoclinic 
orbit (reconnection of a saddle type critical point to itself \cite{js99,strogatz01,chicone06}). We exclude such flows 
in our work and concentrate only on the cases in which we obtain three real solutions for stationary flow. Such configuration 
is characterised by two saddle type critical points with a centre type critical lying between them. One of the saddle type points 
is formed very close to the gravitational horizon in a region with high space-time curvature and is termed as the inner 
critical point while the other is formed far away where the space-time is asymptotically flat and is named as the outer 
critical point. The middle point being a centre type solution does not allow physical flows to occur through it. Depending on the 
values of $\left[{\cal E},\lambda,\gamma\right]$, the inner critical points can form extremely close to the event horizon (even closer than 
the innermost stable circular orbit) and the outer critical points may form at extremely large distances from it. \\

Based on the relative magnitudes of the entropy accretion rate $\dot {\Xi}$ at the inner and the outer critical points, 
the parameter space is categorised into two topologically different subspaces. When $\dot {\Xi}_{inner}>\dot {Xi}_{outer}$, 
a homoclinic orbit is formed through the inner critical point, while for $\dot {\Xi}_{inner}<\dot {Xi}_{outer}$, it is formed 
through the outer critical point. The former represents multitransonic accretion with two saddle type points and a centre type 
critical point in the middle, whereas the latter represents the wind solutions. Since a physical flow cannot occur through centre type 
critical points, the configuration essentially implies transonic flow through the outer and inner critical points. However, it is 
quite intuitive that a smooth continuous flow cannot have multiple transonicity as the flow profile does not provide any smooth trajectory 
for the flow to become subsonic once it has crossed the outer sonic point. It can be realized only when there exists a physically 
allowed discontinuous shock transition to connect two smooth solutions passing separately through the two saddle type critical points. 
Thus in general, multi-critical solutions and multi-transonic solutions are not the same. Real physical multi-transonic flow occurs 
only when the respective criteria for energy preserving shocks (for polytropic accretion) or temperature preserving shocks (for isothermal 
accretion) are satisfied. This will be elaborated in the subsequent sections. \\

The space gradient of the advective flow velocity on the acoustic 
horizon can be computed as 
\begin{equation}
\left(\frac{\text{du}}{\text{dr}}\right)_{r_c}=\left[-\frac{\alpha_1}{2\Gamma_1}\overset{+}{-}\frac{\sqrt{\alpha_1 ^2-4\Gamma_1\beta_1 }}{2\Gamma_1}\right]_{r_c}
\label{anal53}
\end{equation}
where the negative value corresponds to the accretion solution, whereas the positive 
value corresponds to the wind solution. 
$\alpha_1$, $\beta_1$ and $\Gamma_1$ in eq. (\ref{anal53}) have the values
\begin{eqnarray}
\alpha_1 = \left. \frac{2 c_s \left(\gamma -1-c_s^2\right) \left(r-1\right)}{\left(1-c_s^2\right) r \left(r-2\right)}\right|_{r_c}, \nonumber \\
\beta_1 = \left. \frac{\beta'}{(-2+r)^2 r^2 \left(r^3-(-2+r) \lambda ^2\right)^2}\right|_{r_c}, \nonumber \\
\Gamma_1 = \left. \frac{\gamma -3u^2+1}{\left(1-u^2\right){}^2}\right|_{r_c}
\label{anal54}
\end{eqnarray}
where
\begin{flushleft}
$\beta'=[\gamma  \lambda ^4 c_s{}^2 r^4-6 \gamma  \lambda ^4 c_s{}^2 r^3+13 \gamma  \lambda ^4 c_s{}^2 r^2-12 \gamma  \lambda ^4 c_s{}^2 r+4 \gamma  \lambda ^4 c_s{}^2-2 \gamma  \lambda ^2 c_s{}^2 r^6+8 \gamma  \lambda ^2 c_s{}^2 r^5-10 \gamma  \lambda ^2 c_s{}^2 r^4+4 \gamma  \lambda ^2 c_s{}^2 r^3+\gamma  c_s{}^2 r^8-2 \gamma  c_s{}^2 r^7+\gamma  c_s{}^2 r^6-\lambda ^4 r^4-\lambda ^4 c_s{}^4 r^4+8 \lambda ^4 r^3+6 \lambda ^4 c_s{}^4 r^3-24 \lambda ^4 r^2-13 \lambda ^4 c_s{}^4 r^2+\lambda ^4 c_s{}^2 r^2+32 \lambda ^4 r+12 \lambda ^4 c_s{}^4 r-4 \lambda ^4 c_s{}^2 r-4 \lambda ^4 c_s{}^4+4 \lambda ^4 c_s{}^2+2 \lambda ^2 c_s{}^4 r^6+3 \lambda ^2 r^6-8 \lambda ^2 c_s{}^4 r^5-20 \lambda ^2 r^5+10 \lambda ^2 c_s{}^4 r^4+48 \lambda ^2 r^4-2 \lambda ^2 c_s{}^2 r^4-4 \lambda ^2 c_s{}^4 r^3-40 \lambda ^2 r^3+4 \lambda ^2 c_s{}^2 r^3-c_s{}^4 r^8+2 c_s{}^4 r^7-2 r^7-c_s{}^4 r^6+c_s{}^2 r^6+2 r^6-16 \lambda ^4]_{r_c}$
\end{flushleft} 

The critical acoustic velocity gradient
$\left(dc_s/dr\right)_{\rm r=r_c}$ can be obtained by
substituting the value of $\left(\frac{du}{dr}\right)_{\rm r=r_c}$
in eq. (\ref{anal46}).

In order to generate the Mach number vs radial distance plot, one needs to simultaneously integrate 
equations (\ref{anal46} -- \ref{anal47}) for a 
given set of $\left[{\cal E},\lambda,\gamma\right]$.
$\left(\frac{du}{dr}\right)_{\rm r=r_c}$ and $\left(dc_s/dr\right)_{\rm r=r_c}$ are numerically iterated 
using fourth order Runge - Kutta 
method \cite{ptvf07}. Detailed elaboration of the integration scheme 
with representative trajectory plots may be found 
in \cite{das07arxiv,dc12na,pmdc12cqg}.

\subsubsection{Quasi-spherical flow}
\noindent
The expressions for the mass and the entropy accretion rate are
\begin{equation}
\dot{M}=\Lambda \rho \frac{u\sqrt{1-\frac{2}{r}}}{\sqrt{1-u^2}} r^2
\label{anal57}
\end{equation}
and 
\begin{equation}
\dot{\Xi} = \Lambda_{adia} \frac{u\sqrt{1-\frac{2}{r}}}{\sqrt{1-u^2}} r^2 c_s^{\frac{2}{\gamma -1}}(\frac{\gamma -1}{\gamma - (1+c_s^2)})^\frac{1}{\gamma -1}
\label{anal58}
\end{equation}
respectively, where, $\Lambda$ is the geometric solid angle factor. 
The space gradient of the sound velocity and the 
advective velocity are given by,
\begin{equation}
\frac{dc_s}{dr}=-\frac{\gamma -1}{2}\frac{(\frac{1}{u}+\frac{u}{1-u^2})\frac{\text{du}}{\text{dr}}+
\left\{\frac{2}{r}+\frac{1}{r^2(1-\frac{2}{r})}\right\}}{\left\{\frac{c_s}{\gamma -(1+c_s{}^2)}+\frac{1}{c_s}\right\}}
\label{anal59}
\end{equation}
and 
\begin{equation}
\frac{du}{dr} = -\frac{c_s{}^2(\frac{2}{r}+\frac{1}{r^2(1-\frac{2}{r})})+f_2(r,\lambda )}{\frac{u}{1-u^2}(c_s{}^2-1)+
\frac{c_s{}^2}{u}} 
\label{anal60}
\end{equation}
respectively. 
The critical point conditions are derived as
\begin{equation}
\left[u=c_s\right]_{r_c}=\sqrt{\frac{f_2(r_c,\lambda )}{\frac{2}{r_c}+\frac{1}{r_c{}^2(1-\frac{2}{r_c})}}}
\label{anal61}
\end{equation}
Thus, the critical points and the sonic points turn out to be numerically same for quasi-spherical flow.

The critical space gradients of $u$ are obtained as
\begin{equation}
\left(\frac{\text{du}}{\text{dr}}\right)_c=
\left[-\frac{\alpha_2}{2\Gamma_2}\overset{+}{-}\frac{\sqrt{\alpha_2 ^2-4\Gamma_2\beta_2 }}{2\Gamma_2}\right]_{r_c}
\label{anal63}
\end{equation}
where 
\begin{eqnarray}
\alpha_2 = \left.\frac{2 c_s \left(\gamma -1-c_s^2\right) \left(2 r-3\right)}{\left(1-c_s^2\right) r \left(r-2\right)}\right|_{r_c}, \nonumber \\
\beta_2 = \left.\frac{\beta''}{(-2+r)^2 r^2 \left(r^3-(-2+r) \lambda ^2\right)^2}\right|_{r_c}, \nonumber \\
\Gamma_2 = \left.\frac{\gamma -3u^2+1}{\left(1-u^2\right){}^2}\right|_{r_c}
\label{anal64}
\end{eqnarray}
where
\begin{flushleft}
$\beta''=[-36 c^4 \lambda ^4-4 c^4 \lambda ^4 r_c^4+28 c^4 \lambda ^4 r_c^3-73 c^4 \lambda ^4 r_c^2+84 c^4 \lambda ^4 r_c+8 c^4 \lambda ^2 r_c^6-40 c^4 \lambda ^2 r_c^5+66 c^4 \lambda ^2 r_c^4-36 c^4 \lambda ^2 r_c^3-4 c^4 r_c^8+12 c^4 r_c^7-9 c^4 r_c^6+36 c^2 \gamma  \lambda ^4-12 c^2 \lambda ^4+4 c^2 \gamma  \lambda ^4 r_c^4-28 c^2 \gamma  \lambda ^4 r_c^3+73 c^2 \gamma  \lambda ^4 r_c^2-84 c^2 \gamma  \lambda ^4 r_c-8 c^2 \gamma  \lambda ^2 r_c^6+40 c^2 \gamma  \lambda ^2 r_c^5-66 c^2 \gamma  \lambda ^2 r_c^4+36 c^2 \gamma  \lambda ^2 r_c^3+4 c^2 \gamma  r_c^8-12 c^2 \gamma  r_c^7+9 c^2 \gamma  r_c^6-2 c^2 \lambda ^4 r_c^4+14 c^2 \lambda ^4 r_c^3-35 c^2 \lambda ^4 r_c^2+36 c^2 \lambda ^4 r_c+4 c^2 \lambda ^2 r_c^6-20 c^2 \lambda ^2 r_c^5+30 c^2 \lambda ^2 r_c^4-12 c^2 \lambda ^2 r_c^3-2 c^2 r_c^8+6 c^2 r_c^7-3 c^2 r_c^6-\lambda ^4 r_c^4+8 \lambda ^4 r_c^3-24 \lambda ^4 r_c^2+32 \lambda ^4 r_c+3 \lambda ^2 r_c^6-20 \lambda ^2 r_c^5+48 \lambda ^2 r_c^4-40 \lambda ^2 r_c^3-2 r_c^7+2 r_c^6-16 \lambda ^4]_{r_c}$
\end{flushleft}

\subsubsection{Flow in hydrostatic equilibrium in the vertical direction}
\label{vertical-equilibrium-polytropic-flow}
\noindent
Half thickness of the disc at a given radial distance is derived by modifying the 
expression for the flow thickness provided in \cite{alp97apj}
\begin{equation}
H(r) = \frac{r^2c_s}{\lambda }\sqrt{\frac{2(1-u^2)(1-\frac{\lambda ^2}{r^2}(1-\frac{2}{r}))(\gamma -1)}{\gamma (1-\frac{2}{r})(\gamma -(1+c_s{}^2))}}
\label{anal64}
\end{equation}
The mass and entropy accretion rate are thus obtained as
\begin{equation}
\dot{M}=4\pi \rho \frac{u\sqrt{1-\frac{2}{r}}}{\sqrt{1-u^2}} \frac{r^4c_s}{\lambda }\sqrt{\frac{2(1-u^2)(1-\frac{\lambda ^2}{r^2}(1-\frac{2}{r}))(\gamma -1)}{\gamma (1-\frac{2}{r})(\gamma -(1+c_s{}^2))}}
\label{anal65}
\end{equation}
and
\begin{equation}
\dot{\Xi}=\sqrt{\frac{2}{\gamma }}\left[\frac{\gamma -1}{\gamma -(1+c_s{}^2)}\right]{}^{\frac{\gamma +1}{2(\gamma -1)}}\frac{c_s{}^{\frac{\gamma +1}{\gamma -1}}}{\lambda }\sqrt{1-\frac{\lambda ^2}{r^2}(1-\frac{2}{r})} \left( 4 \pi u r^3 \right)
\label{anal66}
\end{equation}
Acoustic and advective velocity gradients are given by,
\begin{equation}
\frac{\text{dc}_s}{\text{dr}} = \frac{-c_s\left\{\gamma -(1+c_s{}^2)\right\}}{\gamma +1}\left[\frac{1}{u}\frac{\text{du}}{\text{dr}}+f_1(r,\lambda )\right]
\label{anal67}
\end{equation}
and
\begin{equation}
\frac{du}{dr} = \frac{\frac{2c_s{}^2}{\gamma +1}f_1(r,\lambda )-f_2(r,\lambda )}{\frac{u}{1-u^2}-\frac{2c_s{}^2}{(\gamma +1)u}} 
\label{anal68}
\end{equation}
 
The critical point condition appears to be
\begin{equation}
\left[u=\sqrt{\frac{1}{1+(\frac{\gamma +1}{2})(\frac{1}{c_s{}^2})}}\right]_{r_c}
=\sqrt{\frac{f_2(r_c,\lambda )}{f_1(r_c,\lambda )+f_2(r_c,\lambda)}}
\label{anal69}
\end{equation}

In this case, it is observed from eq. (\ref{anal69}) 
that $\left[u{\ne}{c_s}\right]_{\rm r=r_c}$. Mach number at the critical point can be easily obtained as 
\begin{equation}
M_c=
\sqrt{
\left({\frac{2}{\gamma+1}}\right)
\frac
{{f_{1}}(r_c,\lambda)}
{{{f_{1}}(r_c,\lambda)}+{{f_{2}}(r_c,\lambda)}}
}
\label{anal69a}
\end{equation}
This value is generally less than $1$. 
For certain sets of $\left[{\cal E},\lambda,\gamma\right]$,
the critical and the sonic points may be separated by
hundreds of gravitational radii. \\

The space gradient of advective velocity can be derived as
\begin{equation}
\left(\frac{\text{du}}{\text{dr}}\right)_c=\left[-\frac{\alpha_3}{2\Gamma_3}\overset{+}{-}\frac{\sqrt{\alpha_3 ^2-4\Gamma_3\beta_3 }}{2\Gamma_3}\right]_{r_c}
\label{anal71}
\end{equation}
where negative sign stands for accretion and where,
\begin{eqnarray}
\alpha_3 = \left.\frac{8 c_s^2 \left(\gamma -1-c_s^2\right) \left(-2 \lambda ^2 r+3 r^3+3 \lambda ^2\right)}{(\gamma +1)^2 r u_s \left(-\lambda ^2 r+r^3+2 \lambda ^2\right)}\right|_{r_c}, \nonumber \\
\beta_3 = \left.\frac{-\beta'''}{(\gamma +1)^2 \left(r-2\right){}^2 r^2 \left(-\lambda ^2 r+r^3+2 \lambda ^2\right){}^2}\right|_{r_c}, \nonumber \\
\Gamma_3 = \left.-\frac{4 c_s{}^4}{(\gamma +1)^2 u^2}+\frac{2 (3 \gamma -1) c_s{}^2}{(\gamma +1)^2 u^2}+\frac{u^2+1}{\left(u^2-1\right){}^2}\right|_{r_c}
\label{anal72}
\end{eqnarray}
with
\begin{flushleft}
$\beta'''=[-36 c_s^4 r^8-30 c_s^2 r^8+42 c_s^2 \gamma  r^8+144 c_s^4 r^7+120 c_s^2 r^7-2 \gamma ^2 r^7-168 c_s^2 \gamma  r^7-4 \gamma  r^7-2 r^7-144 c_s^4 r^6-120 c_s^2 r^6+2 \gamma ^2 r^6+48 c_s^4 \lambda ^2 r^6+42 c_s^2 \lambda ^2 r^6+3 \gamma ^2 \lambda ^2 r^6-54 c_s^2 \gamma  \lambda ^2 r^6+6 \gamma  \lambda ^2 r^6+3 \lambda ^2 r^6+168 c_s^2 \gamma  r^6+4 \gamma  r^6+2 r^6-264 c_s^4 \lambda ^2 r^5-240 c_s^2 \lambda ^2 r^5-20 \gamma ^2 \lambda ^2 r^5+288 c_s^2 \gamma  \lambda ^2 r^5-40 \gamma  \lambda ^2 r^5-20 \lambda ^2 r^5-16 c_s^4 \lambda ^4 r^4-12 c_s^2 \lambda ^4 r^4-\gamma ^2 \lambda ^4 r^4+20 c_s^2 \gamma  \lambda ^4 r^4-2 \gamma  \lambda ^4 r^4-\lambda ^4 r^4+480 c_s^4 \lambda ^2 r^4+456 c_s^2 \lambda ^2 r^4+48 \gamma ^2 \lambda ^2 r^4-504 c_s^2 \gamma  \lambda ^2 r^4+96 \gamma  \lambda ^2 r^4+48 \lambda ^2 r^4+112 c_s^4 \lambda ^4 r^3+84 c_s^2 \lambda ^4 r^3+8 \gamma ^2 \lambda ^4 r^3-140 c_s^2 \gamma  \lambda ^4 r^3+16 \gamma  \lambda ^4 r^3+8 \lambda ^4 r^3-288 c_s^4 \lambda ^2 r^3-288 c_s^2 \lambda ^2 r^3-40 \gamma ^2 \lambda ^2 r^3+288 c_s^2 \gamma  \lambda ^2 r^3-80 \gamma  \lambda ^2 r^3-40 \lambda ^2 r^3-292 c_s^4 \lambda ^4 r^2-216 c_s^2 \lambda ^4 r^2-24 \gamma ^2 \lambda ^4 r^2+368 c_s^2 \gamma  \lambda ^4 r^2-48 \gamma  \lambda ^4 r^2-24 \lambda ^4 r^2+336 c_s^4 \lambda ^4 r+240 c_s^2 \lambda ^4 r+32 \gamma ^2 \lambda ^4 r-432 c_s^2 \gamma  \lambda ^4 r+64 \gamma  \lambda ^4 r+32 \lambda ^4 r-144 c_s^4 \lambda ^4-96 c_s^2 \lambda ^4-16 \gamma ^2 \lambda ^4+192 c_s^2 \gamma  \lambda ^4-32 \gamma  \lambda ^4-16 \lambda ^4]_{r_c}$
\end{flushleft}
 
\noindent
For a given set of initial boundary conditions, 
$\left[u,c_s,du/dr,dc_s/dr\right]_{\rm r_c}$ are used as initial 
values to perform simultaneous numerical integration of eq. (\ref{anal67} -- \ref{anal68}) 
upto the value of $r$ where $u=c_s$.
Values of $\left(du/dr\right)$ and $\left(dc_s/dr\right)$ are 
evaluated at that $r$ ($=r_h$) and then the corresponding set of 
$\left[u,c_s,du/dr,dc_s/dr\right]_{\rm r_s}$ is used to 
evaluate other relevant quantities at the sonic point.

\subsection{Isothermal accretion}
\noindent
For isothermal flow, integration of the stationary general relativistic Euler equation 
leads to the following conserved quantity,
\begin{equation}
\frac{r^2(r-2)}{(r^3-(r-2) \lambda ^2) (1-u^2)} \rho^{2c_s^2}=\text{constant}=\xi
\label{anal73}
\end{equation}
The isothermal sound speed is constant and independent of flow position. It is given by the 
Clapeyron-Mendeleev  equation (\cite{gibbs02,bazarov64})
\begin{equation}
c_s=\sqrt{\frac{k_B}{\mu m_H} T}
\label{anal74}
\end{equation}
where $k_B$ is the Boltzmann's constant, $m_H \approx m_p$ is the mass of hydrogen atom, 
and $\mu$ is the mean molecular weight.

\subsubsection{Flow with constant height H}
\noindent
The mass accretion rate is given by
\begin{equation}
\dot M= 2 \pi \rho \frac{u\sqrt{1-\frac{2}{r}}}{\sqrt{1-u^2}} r h
\label{anal75}
\end{equation}
where $H$ signifies constant height of the disc.
Differentiating eq. (\ref{anal73}) and eq. (\ref{anal75}), the advective velocity gradient is obtained as
\begin{equation}
\frac{du}{dr}=\frac{\left(2r^3-2 (r-2)^2 \lambda^2+(1-r) \left(2r^3+4 \lambda^2-2 r \lambda^2\right) c_s^2\right)u(u^2-1)}{(2 -r) r \left(-2r^3-4 \lambda^2+2 r \lambda^2\right) \left(u^2-c_s^2\right)}
\label{anal76}
\end{equation}
and the critical point conditions are derived as
\begin{equation}
\left[u^2=c_s^2=\frac{- r^3+ (r-2)^2 \lambda ^2}{r^3-r^4+(r-2)(r-1)\lambda ^2}\right]_{r_c}
\label{anal77}
\end{equation}
Thus, the critical and the sonic points are found to coincide. 
Since $c_s\propto{T^\frac{1}{2}}$, for a given set of $\left[T,\lambda\right]$ 
the following polynomial of fourth degree can be solved {\it analytically} to calculate the critical points $r_c$. 
\begin{eqnarray}
2 c_s^2 r^4 - 2 \left(1+c_s^2\right)r^3 - 2\lambda ^2\left(c_s^2-1\right)r^2 - 2\text{$\lambda $}^2\left(4-3c_s^2\right)r \nonumber \\
- 4 \lambda ^2\left(c_s^2-2\right)\bigg\rvert_{r_c} = 0
\label{anal78}
\end{eqnarray}
The gradient of advective velocity at the critical point can then be obtained as
\begin{equation}
\left(\frac{\text{du}}{\text{dr}}\right)_c=
\left[\frac{\alpha_1^{\rm iso}}{2\Gamma_1^{\rm iso}}\overset{+}{-}\frac{\sqrt{{\alpha_1^{iso}}^2+
4\beta_1^{\rm iso}\Gamma_1^{iso}}}{2\Gamma_1^{\rm iso} }\right]_{r_c}
\label{anal79}
\end{equation}
where the $-ve$ sign signifies the accretion solution, and where,
\begin{eqnarray}
\alpha_1^{\rm iso} = -(3 c_s{}^2-1)(-1+c_s{}^2(r_c-1))r{}^3 \nonumber \\
+(3 c_s{}^2-1)(2+c_s{}^2(r-1)-r)(r-2)\lambda ^2 \bigg\rvert_{r_c}, \nonumber \\
\beta_1^{\rm iso}= c_s(1-c_s{}^2)(-3+c_s{}^2(4r-3))r{}^2 \nonumber \\
+c_s(1-c_s{}^2)(c_s{}^2(3-2r)+2(r-2))\lambda ^2\bigg\rvert_{r_c}, \nonumber \\
\Gamma_1^{\rm iso}=2 r (r-2) (r^3 - (r-2) \lambda ^2) c_s\bigg\rvert_{r_c}
\label{anal80}
\end{eqnarray}

\subsubsection{Quasi-spherical Flow}
\noindent
The mass accretion rate is given by
\begin{equation}
\dot M= \Lambda_{iso} \rho \frac{u\sqrt{1-\frac{2}{r}}}{\sqrt{1-u^2}} r^2
\label{anal81}
\end{equation}
$\Lambda$ is the geometric solid angle factor for the flow.
Differentiating eq. (\ref{anal73}) and eq. (\ref{anal81}), the advective velocity gradient is derived as
\begin{equation}
\frac{du}{dr}=\frac{\left\{ 2r^3-2(r-2)^2\lambda ^2+(3-2r)\left(2r^3+4 \lambda ^2-2r \lambda ^2\right)c_s^2\right\}u(u^2-1)}{(2-r)r\left(-2r^3-4 \lambda ^2+2r \lambda ^2\right)(u^2-c_s^2)}
\label{anal82}
\end{equation}
and the critical point conditions become
\begin{equation}
\left[u^2=c_s^2=\frac{- r^3+ (r-2)^2 \lambda ^2}{3 r^3-2 r^4+6 \lambda ^2-7 r \lambda ^2+2 r^2 \lambda ^2} \right]_{r_c}
\label{anal83}
\end{equation}
The critical and the sonic points are found to coincide in these kind of flows as well. 
Using a specific set of astrophysically relevant values for $\left[T,\lambda\right]$ 
the following fourth degree polynomial can be {\it analytically} solved to obtain the critical points $r_c$.
\begin{eqnarray}
4 c_s^2 r^4-2\left(3c_s^2+1\right)r^3-2\lambda ^2\left(2c_s^2-1\right)r^2+2\text{$\lambda $}^2\left(7c_s^2-4\right)r \nonumber \\
-4\lambda ^2\left(3c_s^2-2\right)\bigg\rvert_{r_c} = 0
\label{anal84}
\end{eqnarray}
The advective velocity gradient at the critical point (and hence the sonic point) is obtained as,
\begin{equation}
\left(\frac{\text{du}}{\text{dr}}\right)_c=
\left[\frac{\alpha_2^{\rm iso}}{2\Gamma_1^{\rm iso}}\overset{+}{-}\frac{\sqrt{{\alpha_2^{iso}}^2+
4\beta_2^{\rm iso}\Gamma_1^{iso}}}{2\Gamma_1^{\rm iso} }\right]_{r_c}
\label{anal85}
\end{equation}
where,
\begin{eqnarray}
\alpha_2^{\rm iso} = -(3 c_s{}^2-1)(-1+c_s{}^2(2r_c-3))r^3 \nonumber \\
+(3 c_s{}^2-1)(2+c_s{}^2(2r-3)-r)(r-2)\lambda ^2 \bigg\rvert_{r_c}, \nonumber \\
\beta_2^{\rm iso}= c_s(1-c_s{}^2)(-3+c_s{}^2(8r-9))r^2 \nonumber \\
+c_s(1-c_s{}^2)(c_s{}^2(7-4r)+2(r-2))\lambda ^2\bigg\rvert_{r_c}
\label{anal86}
\end{eqnarray}

\subsubsection{Flow in hydrostatic equilibrium in the vertical direction}
\label{vertical-equilibrium-isothermal-flow}
\noindent
Half thickness of the disc is computed as
\begin{equation}
H(r)^{\rm iso}=\frac{rc_s\sqrt{2(r^3-(r-2 ) \lambda ^2) (1-u^2)}}{\lambda\sqrt{r-2 }}
\label{anal87}
\end{equation}
and the mass accretion rate follows to be
\begin{equation}
\dot M=4\pi \rho \frac{r^2 u c_s}{\lambda} \sqrt{2(r^3-(r-2 ) \lambda ^2)}
\label{anal88}
\end{equation}
The advective velocity gradient is given by 
\begin{equation}
\frac{du}{dr}=\frac{\left[r^3-(r-2)^2 \lambda ^2 +(2 -r) (4 r^3+5 \lambda ^2-3 r \lambda ^2 ) c_s^2 \right] u (u^2-1)}{\frac{1}{2} r (r-2 ) \left(-2r^3-4 \lambda ^2+2 r \lambda ^2 \right) \left[c_s^2-\left(1+c_s^2\right) u^2\right]}
\label{anal89}
\end{equation}
leading to the following critical point condition
\begin{equation}
\left[u^2=\frac{c_s^2}{1+c_s^2}=\frac{- r^3+(r-2)^2 \lambda ^2}{8 r^3-4 r^4+10 \lambda ^2-11 r \lambda ^2+3 r^2 \lambda ^2}\right]_{r_c}
\label{anal90}
\end{equation}
The critical and the sonic points are thus found to be different for such flow geometry. 
As observed in section \ref{vertical-equilibrium-polytropic-flow}, the same is also true 
in the case of vertical equilibrium discs for general relativistic 
isothermal flow in the Schwarzschild metric. However, under the influence of the Newtonian or the pseudo-Schwarzschild black hole
potentials for a similar flow geometry, it had been previously observed that the sonic surfaces and the critical surfaces 
are isomorphic in general(\cite{nard12na}). Hence, for relativistic flow, it is required to integrate eq. (\ref{anal89}) 
in order to find out the exact location of the acoustic horizon $r_h$. 

From eqn. (\ref{anal90}) it is obtained that
\begin{equation} 
M_c=\left[\sqrt{\frac{1}{1+c_s^2}}\right]_{\rm r_c}
\label{anal90a}
\end{equation}
where $M_c$ is the Mach number at the critical point. 
Hence a contrast is observed such non-isomorphism in the case of polytropic flow. 
For isothermal background flow, the value of Mach number is the same irrespective of the location of the 
corresponding critical point due to the constant sound speed. Specifying a set of $\left[T,\lambda\right]$ 
fixes the value of $M_c$. Hence the inner and the outer critical points are collinear on the phase portrait. 
However, for polytropic flows, $M_c$ is a highly nonlinear function of $r_c$ for a given set of 
$\left[{\cal E},\lambda,\gamma\right]$, and hence the two saddle type critical points are not necessarily 
collinear. This result has another important implication. An effective sound speed for isothermal flow in vertical equilibrium 
can be defined as $c_s^{\rm eff}=\frac{c_s}{\sqrt{1+c_s^2}}$, which becomes unity at the critical point. Thus if 
a corresponding acoustic geometry is constructed with such an effective sound speed, the critical and sonic points will be 
isomorphic. However, construction of such an acoustic geometry is impossible for the same geometric flow configuration of a 
polytropic fluid. 

The location of the critical point(s) for a given set of $\left[T,\lambda\right]$ can be solved for 
analytically from the following fourth degree polynomial 
\begin{eqnarray}
4 r^4 c_s^2 - r^3 \left(1+8\text{c}_s^2\right) - r^2 \lambda ^2 \left(-1+3c_s^2\right) - {r\lambda}^2\left(4-11c_s^2\right) \nonumber \\
 - 2\lambda ^2\left(-2+5c_s^2\right)\bigg\rvert_{r_c} = 0
\label{anal91}
\end{eqnarray} 
The critical gradient of velocity is thus obtained as
\begin{equation}
\left(\frac{\text{du}}{\text{dr}}\right)_c=
\left[\frac{\alpha_3^{\rm iso}}{2\Gamma_3^{\rm iso}}\overset{+}{-}\frac{\sqrt{{\alpha_3^{\rm iso}}^2
+4\beta_3^{\rm iso}\Gamma_3^{\rm iso}}}{2\Gamma_3^{\rm iso} }\right]_{r_c}
\label{anal92}
\end{equation}
where,
\begin{eqnarray}
\alpha_3^{\rm iso}= -\left(\frac{3c_s{}^2}{1+c_s{}^2}-1\right)(-1+4c_s{}^2(r-2))r{}^3 \nonumber \\
-\left(\frac{3c_s{}^2}{1+c_s{}^2}-1\right)(2+c_s{}^2(3r-5)-r)(r-2)\lambda ^2\bigg\rvert_{r_c}, \nonumber \\
\beta_3^{\rm iso}= \left(\frac{c_s}{(1+c_s{}^2){}^{\frac{3}{2}}}\right)(-3+8c_s{}^2(2r-3))r{}^2 \nonumber \\
+\left(\frac{c_s}{(1+c_s{}^2){}^{\frac{3}{2}}}\right)(c_s{}^2(11-6r)+2(r-2))\lambda ^2\bigg\rvert_{r_c}, \nonumber \\
\Gamma_3^{\rm iso}=2 r(r-2) (r{}^3 - (r-2) \lambda ^2) c_s\sqrt{1+c_s{}^2}\bigg\rvert_{r_c}
\label{anal93}
\end{eqnarray}

For a given $\left[T,\lambda\right]$, 
$\left[u\right]_{\rm r_c}$ and $\left(du/dr\right)_{\rm r_c}$ are calculated 
as defined by eq. (\ref{anal90}) and eq. (\ref{anal92}) respectively. Then eq. (\ref{anal89}) is integrated from the 
critical point upto the radial distance $r$ ($=r_h$) where the flow becomes transonic. 
$u$ and $du/dr$ are then computed at $r_h$ and used to evaluate other relevant quantities at that point.

As mentioned earlier, for real physical multi-transonic accretion to occur, the flow must pass through 
the inner sonic point. And in order to do so, the flow is required to return to the subsonic regime. However, our system 
so far, does not contain any inherent switch to fulfill this requirement and hence the only way to make it possible would be 
to introduce discontinuities or shocks in the flow. 

In the next section, we describe such discontinuities or {\rm{shocks}} for three different geometric flow 
configurations with both polytropic and isothermal equations of state, and derive corresponding expressions 
for quantities which remain invariant across the shock surface. We investigate the variation of dynamical 
and thermodynamic flow parameters with shock.

\section{Relativistic Rankine-Hugoniot conditions}
Using energy momentum conservation and the continuity equation, the Rankine-Hugoniot conditions 
(\cite{rankine1870ptrsl,hugoniot1887jepa,hugoniot1887jepb}) for a fully general relativistic background flow with 
energy-momentum tensor $T^{\mu\nu}$, 4-velocity $u^{\mu}$, speed of sound $c_s$ and rest mass 
density $\rho$, are given by, \\
\begin{equation}
\left[\left[ \rho u^\mu \right]\right]=0 \text{,and}
\left[\left[T^{\mu\nu} \right]\right]=0,
\label{rh1}
\end{equation}
where $\left[\left[f\right]\right]$ signifies the discontinuity in $f$ across the shock surface.
If $f_-$ and $f_+$ denote the value of $f$ just before and after the shock respectively, then
$\left[\left[f\right]\right]=f_+-f_-$.
For a perfect fluid the energy-momentum tensor is of the form\\
\begin{equation}
T^{\mu\nu}=(p+\epsilon)u^{\mu}u^{\nu}+pg^{\mu\nu},
\label{T}
\end{equation}
where $p$ and $\epsilon$ denote pressure and energy density of the fluid respectively. 
In all further calculations, the above form has been considered in Boyer-Lindquist (\cite{bl67jmp}) 
co-ordinates normalized for $G=M_{BH}=c=1$ and $\theta=\frac{\pi}{2}$. The radial velocity of the 
flow along the direction of the normal $\eta_\mu$ to the shock hypersurface is given by,
\begin{equation}
u^r=\frac{u \Delta^\frac{1}{2}}{r\sqrt{1-u^2}},
\label{ur}
\end{equation}
where $\Delta=r(r-2)$. Hence, the Rankine-Hoguniot conditions are obtained as, \\
\begin{equation}
\left[\left[\rho u^r\right]\right]=0,
\label{rh2}
\end{equation}
\begin{equation}
\left[\left[T_{t\mu}u^\mu\right]\right]=\left[\left[(p+\epsilon)u_tu^r\right]\right]=0,
\label{rh3}
\end{equation}
\begin{equation}
\left[\left[T_{\mu\nu}\eta^\mu\eta^\nu\right]\right]=\left[\left[(p+\epsilon)u^ru^r+p\right]\right]=0.
\label{rh4}
\end{equation}
Here time component of 4-velocity $u^\mu$ is given by $u_t=\sqrt{\frac{\Delta}{B(1-u^2)}}$, 
where $B=r^2-\lambda^2(1-\frac{2}{r})$. 

\section{Shock-invariant quantities ($S_h$)}
\subsection{Polytropic accretion}
For polytropic flow, the stationary shocks are energy preserving. 
Using eqns.(\ref{eqnofstatepoly}),(\ref{enthalpy1}),(\ref{epsilon}),(\ref{csq1}),(\ref{enthalpy3}), 
one can express $\rho$, $p$ and $\epsilon$ in terms of the sound speed $c_s$ as,
\begin{equation}
\rho=K^{-\frac{1}{\gamma-1}}\left({\frac{\gamma-1}{\gamma}}\right)^{\frac{1}{\gamma-1}}\left({\frac{c_s^2}{\gamma-1-c_s^2}}\right)^{\frac{1}{\gamma-1}}
\label{rho}
\end{equation}
\begin{equation}
p=K^{-\frac{1}{\gamma-1}}\left({\frac{\gamma-1}{\gamma}}\right)^{\frac{\gamma}{\gamma-1}}\left({\frac{c_s^2}{\gamma-1-c_s^2}}\right)^{\frac{\gamma}{\gamma-1}}
\label{p}
\end{equation}
\begin{equation}
\epsilon=K^{-\frac{1}{\gamma-1}}\left({\frac{\gamma-1}{\gamma}}\right)^{\frac{1}{\gamma-1}}\left({\frac{c_s^2}{\gamma-1-c_s^2}}\right)^{\frac{1}{\gamma-1}}\left[1+\frac{1}{\gamma}\left(\frac{c_s^2}{\gamma-1-c_s^2}\right)\right]
\label{e}
\end{equation}
So far, geometric configuration of the accretion disc has not played its part. However, on careful 
observation, one can instantly infer that the mass continuity equation, when formulated for a 
specific type of disc, shall contain an additional area factor $\mathcal{A}(r)$ whose functional form will 
depend exclusively on the geometry of the background flow. As mentioned earlier, we are going to 
focus on three distinct flow configurations, viz. discs with constant height H (denoted hereafter 
by 'CH'), quasi-spherical discs tracing a solid angle $\Lambda$ (denoted hereafter by 'CF') and 
discs in hydrostatic equilibrium in the vertical direction with radius dependent height $H(r)$ 
(denoted hereafter by 'VE'). Thus, eq.(\ref{rh2}) will take the from, \\
\begin{equation}
\left[\left[\rho u^r\mathcal{A}(r)\right]\right]=0
\label{rh5}
\end{equation}
where, \\
\begin{equation}
\mathcal{A}_{CH}(r)=2\pi rH,
\label{ach}
\end{equation}
\begin{equation}
\mathcal{A}_{CF}(r)=\Lambda r^2, and
\label{acf}
\end{equation}
\begin{equation}
\mathcal{A}_{VE}(r)=4\pi rH(r).
\label{ave}
\end{equation} 
We now define the \text{\it{shock-invariant quantity}} ($S_h$). Numerical value of $S_h$ remains the same on the 
integral solution passing through the outer sonic point as well as the solution passing through inner sonic point, 
only at the shock location and nowhere else. If $S_h^{in}$ and $S_h^{out}$ are the shock-invariant quantities defined 
on the flow topologies through the inner and outer sonic points respectively, then the values of the radial distance 
for which $S_h^{out}-S_h^{in}=0$ provides the shock location. \\

Substituting eq.(\ref{ach}), (\ref{acf}), (\ref{ave}) in eq.(\ref{rh5}), (\ref{rh4}) and solving 
simultaneously, the \text{\it{shock-invariant quantities}} ($S_h$) for all three geometries are 
derived as,
\begin{equation}
S_h\bigr\rvert_{CH}=\frac{u^2(\gamma\frac{\Delta}{r^2}-c_s^2)+c_s^2}{u\sqrt{1-u^2}\left(\gamma-1-c_s^2\right)}
\label{shch}
\end{equation}
\begin{equation}
S_h\bigr\rvert_{CF}=\frac{u^2(\gamma\frac{\Delta}{r^2}-c_s^2)+c_s^2}{u\sqrt{1-u^2}\left(\gamma-1-c_s^2\right)}
\label{shcf}
\end{equation}
\begin{equation}
S_h\bigr\rvert_{VE}=\frac{\lambda{\Delta}^{\frac{1}{2}}\{u^2(\gamma\frac{\Delta}{r^2}-c_s^2)+c_s^2\}}{uc_s(1-u^2)\sqrt{\gamma-1-c_s^2}}
\label{shve}
\end{equation}
Here, it is required to note that width of shock must be very small. Otherwise, strong temperature 
gradients along the shock region can lead to energy dissipation through the extreme ends.

\subsection{Isothermal accretion}
For isothermal flow, the stationary shocks are temperature preserving.
$\xi$ is conserved across the shock surface. Hence, substituting $\rho$ 
from the mass continuity equation,
\begin{equation}
\dot{M}=\rho u^r\mathcal{A}(r),
\label{masscon}
\end{equation}
a Rankine-Hugoniot condition analogous to eq.(\ref{rh3}) can be obtained as,
\begin{equation}
\left[\left[\xi\right]\right]=\left[\left[u_t^2\{u^r\mathcal{A}(r)\}^{-2c_s^2}\right]\right]=0
\label{xi2}
\end{equation}
Now, since $P=K\rho$ and $c_s^2=\frac{1}{h}\frac{dP}{d\rho}$, where specific enthalpy $h=\frac{p+\epsilon}{\rho}$,
\begin{equation}
\epsilon=\frac{K\rho}{c_s^2}\left(1-c_s^2\right).
\label{epsiloniso}
\end{equation}
Therefore, the shock condition pertaining to momentum conservation (eq.\ref{rh4}) in this 
case becomes,
\begin{equation}
\left[\left[\{1+\frac{u^ru^r}{c_s^2}\}\{u^r\mathcal{A}(r)\}^{-1}\right]\right]=0
\label{momconiso}
\end{equation}
Solving eq.(\ref{xi2}) and eq.(\ref{momconiso}) simultaneously, we derive the \text{\it{shock-invariant quantities}} for three distinct geometries,
\begin{equation}
S_h\bigr\rvert^{iso}_{CH}=\bigr\{\frac{u}{\sqrt{1-u^2}}\bigr\}^{2c_s^2-1}\{u^2\Delta+r^2c_s^2(1-u^2)\},
\label{shchiso}
\end{equation}
\begin{equation}
S_h\bigr\rvert^{iso}_{CF}=\bigr\{\frac{u}{\sqrt{1-u^2}}\bigr\}^{2c_s^2-1}\{u^2\Delta+r^2c_s^2(1-u^2)\},
\label{shcfiso}
\end{equation}
\begin{equation}
S_h\bigr\rvert^{iso}_{VE}=u^{2c_s^2-1}\{u^2\Delta+r^2c_s^2(1-u^2)\}.
\label{shveiso}
\end{equation}

\section{Dependence of shock on flow geometry}
  
\subsection{Polytropic accretion}

In this section, effect of the geometry of an axially symmetric, non self-gravitating, stationary background flow in 
Schwarzschild space-time on the properties of discontinuities or shocks for polytropic accretion onto astrophysical 
black holes is investigated. The parameter space determined by initial boundary conditions is consturcted first. 
Then properties of the shock are depicted for certain regions of $\left[{\cal E},\lambda,\gamma\right]$ for which
multi-transonic accretion with stable shocks can be obtained for all three flow configurations.

\subsubsection{Parameter space classification}

The critical point(s) are 
obtained by specifying suitable values for $\left[{\cal E},\lambda,\gamma\right]$. The 3D parameter space $\left[{\cal E},\lambda,\gamma\right]$ 
has bounds $\left[1<{\cal E}<2,2<\lambda{\le}4,4/3{\le}\gamma{\le}5/3\right]$, which are astrophyscially relevant and of our interest. 
We scan the given parameter space to understand the dependence of multi-criticality on initial boundary conditions. 

In figure 1, the $\left[{\cal E},\lambda,\gamma\right]$ space has been projected on a $\left[{\cal E},\lambda\right]$ plane 
with $\gamma=1.4$. Such $\left[{\cal E},\lambda\right]$
projections may be studied for other values of $\gamma$ as well, lying in the range $4/3{\le}\gamma{\le}5/3$. 

\begin{figure}
\centering
\includegraphics[scale=0.6]{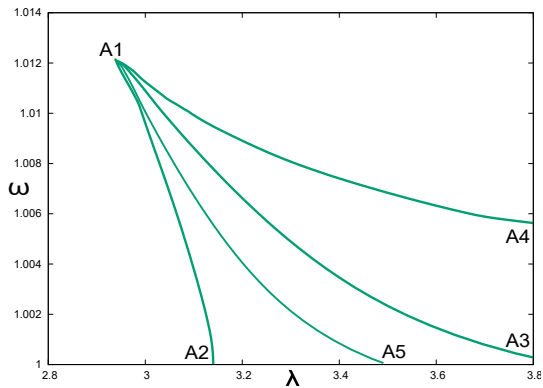}
\caption[]{${\cal E} - \lambda$ plane for quasi-spherical flow geometry for polytropic accretion
for fixed value of $\gamma=1.4$.}
\label{fig1}
\end{figure}

Figure (\ref{fig1}) depicts the ${\cal E} - \lambda$ plane for quasi-spherical disc geometry.
A$_1$A$_2$A$_3$A$_4$ contains three real 
positive roots lying outside event horizon for the corresponding polynomial equation in $r_c$. 
A$_1$A$_2$A$_3$ represents the multi-critical accretion region for which ${\dot {\Xi}}_{\rm inner} > {\dot {\Xi}}_{\rm outer}$. 
Subspace A$_1$A$_5$A$_3$ allows shock formation. This region contains real physical multi-transonic stationary accretion solutions 
where transonic flow passing through the outer sonic point meets transonic flow through the inner sonic point 
by means of a discontinuous energy preserving Rankine-Hugoniot type shock.
Thus, such a shocked multi-transonic solution consists of smooth transonicities at two regular saddle points 
and a discontinuous transonicity at the location of shock.

Region A$_1$A$_3$A$_4$ represents a subset of the parameter space where 
${\dot {\Xi}}_{\rm inner} < {\dot {\Xi}}_{\rm outer}$. 
Here, the incoming flow possesses one saddle type critical point and the background flow
consists of one acoustic horizon at this inner sonic point. 
Multi-critical accretion at the boundary A$_1$A$_3$ separating these two regions is characterized by 
${\dot {\Xi}}_{\rm inner} = {\dot {\Xi}}_{\rm outer}$. Transonic solutions on the boundary 
are completely degenerate and the phase portrait thus formed leads to a heteroclinic orbit 
\footnote{Heteroclinic orbits are the trajectories defined on a phase
portrait which connects two different saddle type critical points. Integral
solution configuration on phase portrait characterized by heteroclinic
orbits are topologically unstable.} which may be unstable and turbulent.

\begin{figure}
\centering
\includegraphics[scale=0.65]{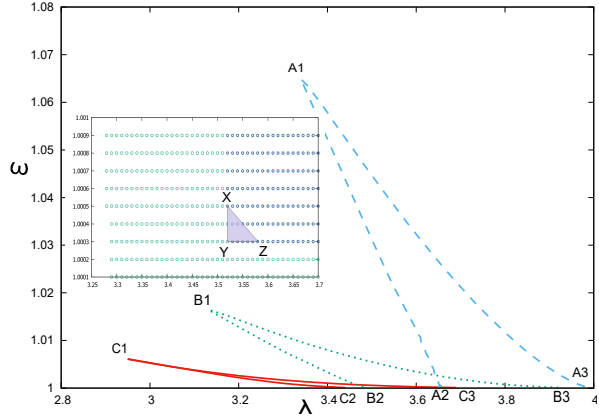}
\caption[]{${\cal E} - \lambda$ plane for three different flow geometries with shock
for fixed value of $\gamma=4/3$. 
Variation of ${\cal E} - \lambda$ branches for flow in hydrostatic equilibrium
along the vertical direction, conical flow and flow with constant thickness
are represented by solid red lines, dashed green lines, and dotted blue lines, respectively.
Inset: Shaded region XYZ shows the overlap of shocked multitransonic accretion solutions 
for all three matter geometries.}
\label{fig2}
\end{figure}

Figure (\ref{fig2}) depicts the same $\left[{\cal E},\lambda\right]$ parameter space diagram with value of $\gamma$ fixed at $4/3$ 
for all the three flow geometry configurations with only those multi-transonic accretion solutions of $r_c$ which 
allow the formation of standing shocks. As described in the previous subsection, these are subspaces of the entire mathematically 
permissible $\left[{\cal E},\lambda\right]$ space providing three real roots of the corresponding polynomial equations in $r_c$ 
outside the event horizon. A$_1$A$_2$A$_3$, B$_1$B$_2$B$_3$ and C$_1$C$_2$C$_3$ shown using dotted blue lines, 
dotted green lines and solid red lines represent flow with constant thickness, quasi-spherical or conical flow and flow in vertical 
hydrostatic equilibrium respectively. It may be observed from the figure that a very small overlap region (region $XYZ$ in the inset) 
of the three wedges does exist, thus ensuring a common astrophysically relevant range of $\left[{\cal E},\lambda\right]$ for the 
given value of $\gamma$ for which true physical multi-transonic accretion can occur. We shall be interested in this common region 
of physical flow solutions while comparing the three different disc configurations for all our future purposes.

\subsubsection{Methodology for obtaining multi-transonic topology with shock}

Using a specific set of $\left[\cal E, \lambda, \gamma\right]$, one first solves the equation for 
$\cal E$ at the critical point to find out the corresponding three critical points, saddle type inner, 
center type middle, and the saddle type outer. The space gradient of the flow velocity as well as the acoustic 
velocity at the saddle type critical point is then obtained. Such $u_{|_{(r=r_{c})}}$, $c_{s_{(r=r_{c})}}$,
$\frac{dc_s}{dr}|_{(r=r_{c})}$ and $\frac{du}{dr}|_{(r=r_{c})}$, serve as the initial value condition for 
performing the numerical integration of the advective velocity gradient using the fourth-order Runge-Kutta 
method. Such integration provides the outer sonic point (located closer to the black hole compared to the 
outer critical point, since the Mach number at the outer critical point is less than unity), the local 
advective velocity, the polytropic sound speed, the Mach number, the fluid density, the disc height, the bulk 
temperature of the flow and any other relevant dynamical and thermodynamic quantity characterizing the flow. 
The corresponding wind solution through the outer sonic point is obtained in the same way, by taking the 
other value of $\frac{du}{dr}|_{(r=r_{c})}$ (note that solution of the equation in 
$\frac{du}{dr}|_{(r=r_{c})}$ which is a quadratic algebraic equation, provides two initial values, one for 
the accretion and the other for the wind). \\

The respective accretion and the wind solutions passing through the inner critical point are obtained 
following exactly the same procedure as has been used to draw the accretion and wind topologies passing 
through the outer critical point. Note, however, that the accretion solution through the inner critical point 
folds back onto the wind solution which is a homoclinic orbit through the inner critical point encompassing 
the center type middle critical point. A physically acceptable transonic solution must be globally consistent, 
i.e. it must connect the radial infinity with the black hole event horizon. Hence, for multi-transonic 
accretion there is no individual existence of physically acceptable accretion/wind solution passing through 
the inner critical (sonic) point, although, depending on the inital boundary conditions, such solution 
may be clubbed with the accretion solution passing through outer critical point, through a standing shock. \\

The set $\left[\cal E, \lambda\right]_{A}$ thus produces doubly degenerate accretion/wind solutions. Such two fold 
degeneracy may be removed by the entropy considerations since the entropy accretion rates for solutions passing 
through the inner critical point and the outer critical point are not generally equal. For any 
$\left[\cal E, \lambda, \gamma\right] \in \left[\cal E, \lambda, \gamma\right]_{A}$ we find that the entropy 
accretion rate $\dot{\Xi}$ evaluated for the complete accretion solution passing through the outer critical point 
is less than that of the rate evaluated for the incomplete accretion/wind solution passing through the inner critical point . 
Since the quantity $\dot{\Xi}$ is a measure of the specific entropy density of the flow, the solution passing 
through the outer critical point will naturally tend to make a transition to its higher entropy counterpart, i.e., 
the globally incomplete accretion solution passing through the inner critical point. Hence, if there existed a 
mechanism for the accretion solution passing through the outer critical point to increase its entropy accretion rate 
exactly by an amount
\begin{equation}
\Delta\dot{\Xi}=\dot{\Xi}(r^{in}_c)-\dot{\Xi}(r^{out}_c),
\end{equation}
there would be a transition to the accretion solution passing through the inner critical point. Such a transition would 
take place at a radial distance somewhere between the radius of the inner sonic point and the radius of the point of 
inflexion of the homoclinic orbit. In this way one would obtain a true multi-transonic accretion solution 
connecting the infinity and the event horizon, ehich includes a part of the accretion solution passing through 
the inner critical, and hence the inner sonic point. One finds that for some specific values of $\left[\cal E, \lambda, \gamma\right]_{A}$, 
a standing Rankine-Hugoniot shock may accomplish this task. A supersonic accretion through the outer sonic point 
(which is obtained by integrating the flow startinf from the outer critical point) can generate entropy through 
such a shock formation and can join the flow passing through the inner sonic point (which is obtained by integratign 
the flow starting from the outer critical point). Below we will carry on a detailed discussion on such shock formation. \\

In the presence of a shock, the flow may have the following profile. A subsonic flow starting from infinity first 
becomes supersonic after crossing the outer sonic point and somewhere in between the outer sonic point and the inner 
sonic point, the shock transition takes place and forces the solution to jump onto the corresponding subsonic branch. 
The hot and dense post-shock subsonic flow produced in this way becomes supersonic again after crossing the inner 
sonic point and ultimately dives supersonically into the black hole. \\

The shock location in multi-transonic accretion is found in the following way. While performing the numerical 
integration along the solution passing through the outer critical point, we calculate the shock invariant 
$S_h$ in addition to $u,c_s$ and $M$. We also calculate $S_h$ while integrating along the solution passing 
through the inner critical point, starting from the inner sonic point upto the point of inflexion of the homoclinic 
orbit. We then determine the radial distance $r_{sh}$, where the numerical values of $S_h$, obtained by 
integrating the two different sectors described above, are equal. Generally, for any value 
of $\left[\cal E, \lambda, \gamma\right]$ allowing shock formation, one finds two formal shock 
locations, one located in between the outer and the middle sonic point, and the other located 
in between the inner and the middle sonic point. The shock strength is different for the inner and for the 
outer shock. Acording to the standard local stability analysis (Yang and Kafatos, 1995), for a 
multi-transonic accretion, one can show that only the shock formed between the middle and outer sonic 
point is stable. Hereafter, whenever we mention the shock location, we always refer to the stable 
shock location only.

\begin{figure}
\centering
\includegraphics[scale=0.4]{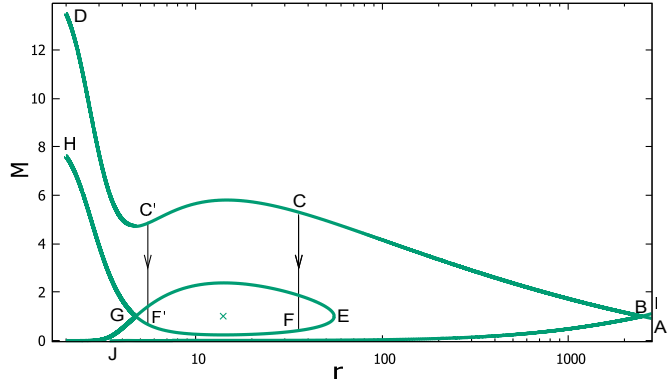}
\caption[]{Topology (Mach number $\left(M\right)$ - $r$ plot) for quasi-spherical flow 
geometry with shock for fixed value of $\gamma=4/3$, $\mathcal{E}=1.0003$ and $\lambda=3.6$.}
\label{fig2a}
\end{figure}

Figure (\ref{fig2a}), depicts the topology of shocked multi-transonic accretion/wind for quasi-spherical or 
conical disc flow geometry. The value of $\gamma$ has been fixed at $\frac{4}{3}$ and those of 
$\cal E$ and $\lambda$ have been chosen from the set which allows formation of standing shock. 
In the figure, the upper branch $ABCC'D$ represents the integral solution from a very large radial distance 
passing through the outer critical point and the outer sonic point (which are equal in this case), contonuing 
till the event horizon, in case there are no discontinuities in the flow. The lower branch $EFF'GH$ actually 
represents two seperate branches of the integral solution- branch $GF'FE$ from the inner critical/sonic 
point (equal in the present case) to the point of inflexion of the homoclinic orbit surrounding 
the middle critical point, and branch $GH$ from the inner critical/sonic point till the event horizon. Now 
if the flow starting from a very large radial distance after passing through the puter sonic point 
does not face any shock or discontinuity, it will continue undisturbed along the upper branch and end up 
falling into the black hole. However such flow will not become subsonic for a second time and hence will 
not be multi-transonic. Therefore, for a multi-transonic flow to occur the flow must jump from from the 
upper branch to the lower branch through a Rankine-Hugoniot type energy preserving shock. Such discontinuous 
jump locations are calculated using procedures discussed in the previous sections. The vertical lines $CF$ and 
$C'F'$ represent the analytically obtained discontinuous jump locations for the depicted flow. As elaborated 
in the preceding paragraph, $CF$ is the required physical stable shock, whereas $C'F'$ is unstable and hence 
is not of our physical interest. Thus, the flow originating from infinity becomes supersonic after 
passing through the outer sonic point, faces a discontinuity at $C$, jumps onto the subsonic lower branch at 
$C'$, becomes supersonic again after passing through the inner sonic point, and finally takes its plunge of 
death into the black hole.

\subsubsection{Dependence of shock related quantities on flow geometry}

\begin{figure}
\centering
\includegraphics[scale=0.4]{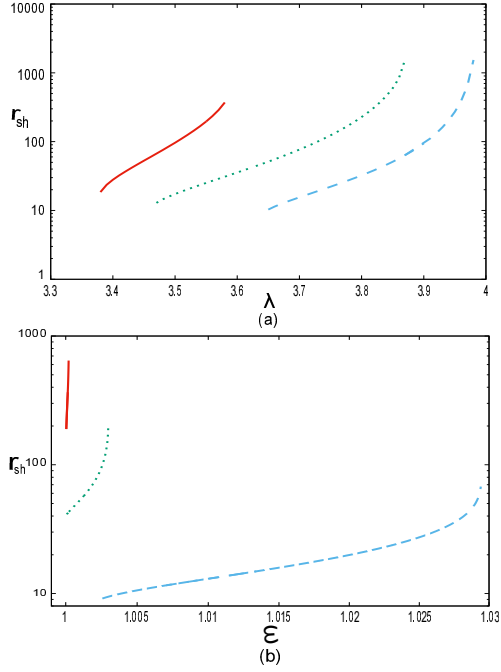}
\caption[]{(a) $r_{sh} - \lambda$ variation for three different flow geometries for polytropic accretion.
$\mathcal{E}=1.0003$ and $\gamma=\frac{4}{3}$ for all three branches. (b)  $r_{sh} - \cal E$ variation for 
three different flow geometries for polytropic accretion. $\lambda=3.6$ and $\gamma=\frac{4}{3}$ for all three branches.
Flow in hydrostatic equilibrium along the vertical direction, conical flow and flow with constant 
thickness are represented by solid red lines, dotted green lines, and dashed blue lines, respectively.}
\label{fig3}
\end{figure}

In figure \ref{fig3}(a) we show the variation of the shock location $r_{sh}$ with $\lambda$. The value of the polytropic index 
$\gamma$ has been fixed at $4/3$. The value of $\cal E$ has been taken to be $1.0003$ such that we get real multi-transonic 
solutions with stable shocks for all the three flow geometries. It is observed that the location of the discontinuity varies over 
a very wide range of the order of $10$s to $1000$s of the Schwarschild radius. We also observe that the distance of the location 
of shock formation from the event horizon increases with increasing specific angular momentum of the flow. 
The correlation of $r_{sh}$ with $\lambda$ is obvious because higher the flow 
angular momentum, the greater the rotational energy content of the flow. As a consequence, the 
strength of the centrifugal barrier which is responsible to break the incoming flow by forming 
a shock will be higher and the location of such a barrier will be farther away from the event horizon. 
A similar trend is observed in figure \ref{fig3}(b) where variation of $r_{sh}$ with the specific energy $\cal E$ 
has been depicted. The value of $\gamma$ remains fixed at $4/3$ and $\lambda$ has been fixed at $3.6$ so 
that real multi-transonic shocked solutions are obtained for all three geometries.

\begin{figure}
\centering
\includegraphics[scale=0.5]{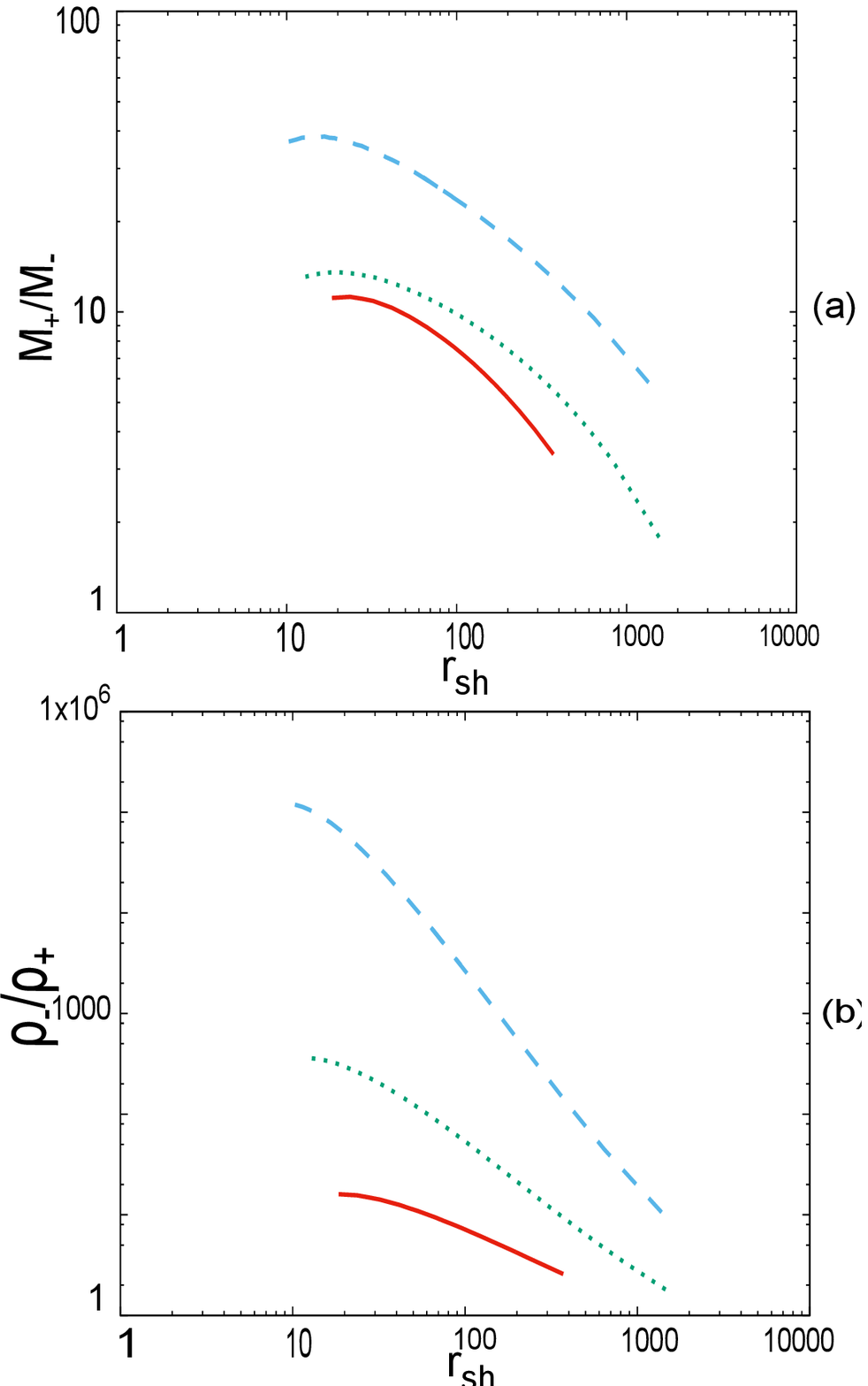}
\includegraphics[scale=0.5]{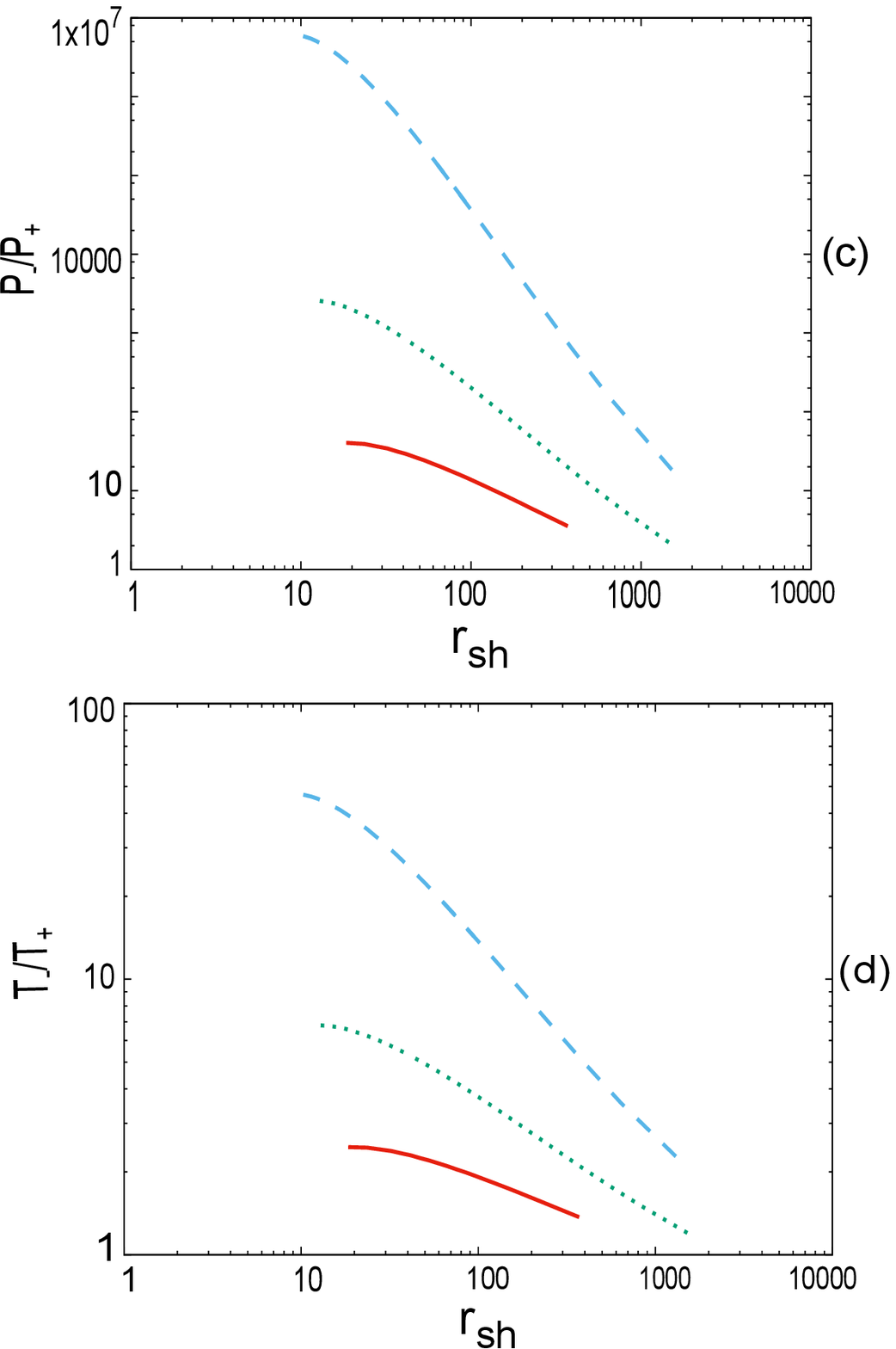}
\caption[]{(a)$M_+/M_-- r_{sh}$, (b)$\rho_-/\rho_+ - r_{sh}$, (c)$P_-/P_+ - r_{sh}$ and (d)$T_-/T_+ - r_{sh}$ 
variations for three different flow geometries for polytropic accretion. $\mathcal{E}=1.0003$ and $\gamma=\frac{4}{3}$ 
for all three branches.
Flow in hydrostatic equilibrium along the vertical direction, conical flow and flow with constant 
thickness are represented by solid red lines, dotted green lines, and dashed blue lines, respectively.}
\label{fig4}
\end{figure}

Figure (\ref{fig4}) depicts the variation of shock strengths in terms of (a) Mach number before and after the shock ($M_+/M_-$), 
(b) density of infalling matter after and before the shock ($\rho_-/\rho_+$), (c) pressure of the infalling matter after and 
before the shock ($P_-/P_+$) and (d) temperature of flow after and before the shock ($T_-/T_+$) with the distance of 
shock location $r_{sh}$ from the event horizon. It is evident 
from the plot that as the shock location shifts closer to the gravitational horizon, the magnitude of jump for the
corresponding quantities rises, thus in turn indicating an increase in strength of the shock. The closer to the 
black hole the shock forms, the higher are the strengths and the entropy enhancement ratio. The ultra relativistic 
flows are supposed to produce the strongest shocks. The reason behind this is also easy to understand. The closer 
to the black hole the shock forms, the higher the available gravitational potential energy must be released, and the 
radial advective velocity required to have a more vigorous shock jump will be larger

\subsection{Isothermal accretion}

First, we construct the $\left[T,\lambda\right]$ parameter space showing the astrophysically relevant range of parameters 
for which real multi-critical solutions can be obtained. Then a specific subset of the entire parameter space is chosen for 
which all three flow geometries will display multi-transonic isothermal accretion with Rankine-Hugoniot type 
discontinuity in the flow.

\subsubsection{Parameter space classification}

\begin{figure}
\centering
\includegraphics[scale=0.7]{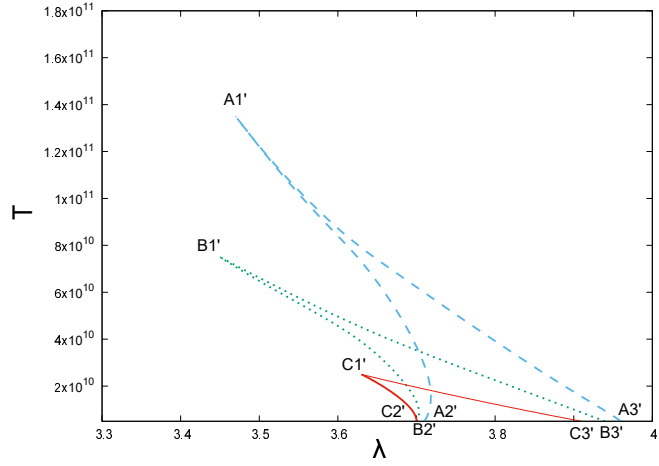}
\caption[]{$T - \lambda$ plane for three different flow geometries with isothermal shock.
$T$ is in units of Kelvin.
Flow in hydrostatic equilibrium along the vertical direction, conical flow and flow with constant 
thickness are represented by solid red lines, dotted green lines, and dashed blue lines, respectively.}
\label{fig5}
\end{figure}

Figure (\ref{fig5}) shows the parameter space division following a similar scheme as introduced in the previous subsection. 
For constant height flow, conical flow and vertical equilibrium flow, 
A1${^\prime}$A2${^\prime}$A3${^\prime}$,
B1${^\prime}$B2${^\prime}$B3${^\prime}$ and 
C1${^\prime}$C2${^\prime}$C3${^\prime}$, respectively 
represent the $\left[T,\lambda\right]$ regions for which 
the algebraic polynomial equations in $r_c$ will 
provide three real physical roots located outside the gravitational horizon, alongwith the added constraint 
that the obtained values of $r_c$ also satisfy the Rankine-Hugoniot criteria, which in this case is ensured by invariance of 
the quantitites $S_h$  given by eqn.(\ref{shchiso}), eqn.(\ref{shcfiso}) and eqn.(\ref{shveiso}).
Similar to the case of polytropic accretion, such shocks provide true multi-transonicity. 
However, unlike polytropic accretion, in the isothermal case the shocks have to be temperature preserving. 
The solutions consist of two acoustic black hole horizons (inner and outer) and 
an acoustic white hole (at the shock). The analysis of this shock has been performed in our present work.

\subsubsection{Dependence of shock related quantitites on flow geometry}

\begin{figure}
\centering
\includegraphics[scale=0.4]{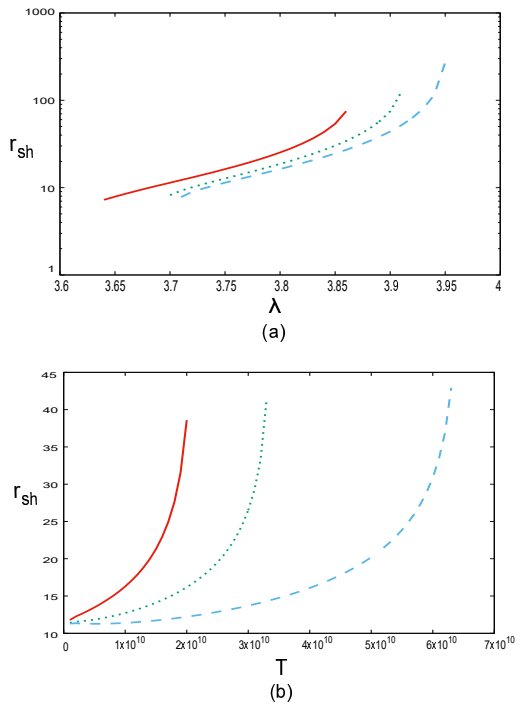}
\caption[]{(a) $r_{sh} - \lambda$ variation for three different flow geometries for isothermal accretion.
$T=0.0001$ (in terms of $10^{10}$ K) for all three branches. 
(b) $r_{sh} - T$ (in terms of Kelvin) variation for three different flow geometries for isothermal accretion.
$\lambda=3.76$ for all three branches.
Flow in hydrostatic equilibrium along the vertical direction, conical flow and flow with constant 
thickness are represented by solid red lines, dotted green lines, and dashed blue lines, respectively.}
\label{fig6}
\end{figure}

Figure \ref{fig6}(a) represents the variation of the shock location $r_{sh}$ with specific angular momentum $\lambda$ 
for all the three flow geometries. The value of $T$ has been chosen from the region of parameter space overlap 
and has been fixed at $0.0001$ (in terms of $10^{10}$ Kelvin). Figure \ref{fig6}(b) depicts the variation of $r_{sh}$ with 
flow temperature $T$, where $\lambda$ has been fixed at the value $3.76$. It is observed that although there is 
a similar rising trend of $r_{sh}$ as was observed in polytropic accretion, however, the variation is over a much 
larger range with $\lambda$ than with $T$.

\begin{figure}
\centering
\includegraphics[scale=0.6]{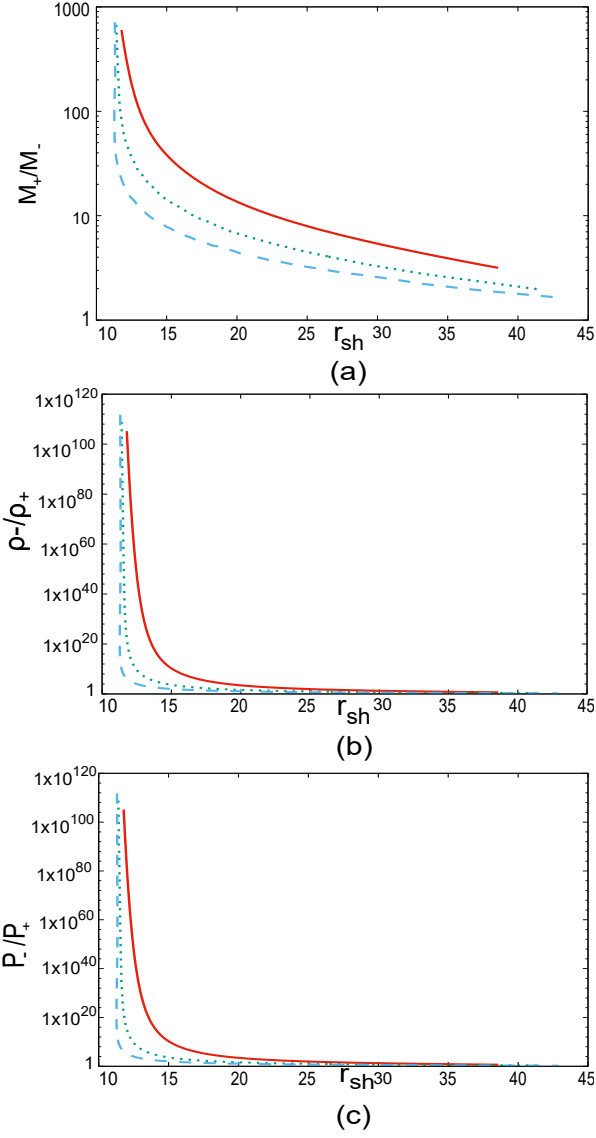}
\caption[]{(a) $M_-/M_+ - r_{sh}$, (b) $\rho_+/\rho_- - r_{sh}$ and (c) $P_+/P_- - r_{sh}$ variation 
for three different flow geometries for isothermal accretion.
$T=0.0001$ (in terms of $10^{10}$ K) for all three branches. 
Flow in hydrostatic equilibrium along the vertical direction, conical flow and flow with constant 
thickness are represented by solid red lines, dotted green lines, and dashed blue lines, respectively.}
\label{fig7}
\end{figure}

In figure (\ref{fig7}), we report the variation of shock strengths in terms of (a) Mach number before and after the shock ($M_-/M_+$), 
(b) density of infalling matter after and before the shock ($\rho_+/\rho_-$), and (c) pressure of the infalling matter after and 
before the shock ($P_+/P_-$) with the distance of 
shock location $r_{sh}$ from the event horizon. Akin to the results obtained for polytropic accretion, here also it is observed 
that as the shock location moves closer to the horizon, the magnitude of jump for the
corresponding quantities rises, thus indicating a stronger shock. It is worth noting that the magnitude of the discontinuity 
in isothermal accretion is far more prominent than that in polytropic accretion. While the Mach number jumps upto a factor of 
1000, the pressure and density of the infalling matter increases all of a sudden by huge orders of magnitude. Since the shock 
is temperature preserving, it has to dissipate an immense quantity of energy, which probably is the reason for peoducing such extremely 
strong shocks.

\section{Concluding remarks}

We demonstrate that physical multitransonic accretion 
with shocks does occur in all flow geometries for a common parameter space of $\left[\cal E,\lambda,\gamma\right]$ for polytropic 
fluid and $\left[T,\lambda\right]$ for isothermal fluid flow. Based on our obtained results a comparision 
between polytropic and isothermal accretion revealed that the effects of discontinuities in the flow are much more prominent 
in the isothermal case. It is also observed that although the shocks are much more prominent in 
isothermal accretion, but a comparision among the different flow geometries depicts that the isothermal shock 
strengths remain of comparable magnitude for all three configurations with slight variations in the rate of change of the 
shock strength related quantitites. \\

For axially symmetric relativistic accretion onto a non-rotating black hole, the geometric 
configuration of the infalling matter influences the accretion dynamics in general. The transonic features of stationary 
integral flow solutions depends on matter geometry, although the no-self gravity of the flow is taken into account 
(and hence no backreaction of the metric). For same set of initial boundary conditions, flow in hydrostatic equilibrium 
in the vertical direction produces the sonic points closest to the horizon whereas the sonic points form at the farthest 
distance for constant height flow. Conical flow in characterised by production of the sonic points lying in between these 
two models. The rate at which the difference of the space gradient of the advective velocity and the adiabatic sound speed, 
i.e. the quantity $\left(\frac{du}{dr}-\frac{dc_s}{dr}\right)$, increases with the radial distance (measured from the horizon) 
is thus minimnum for flow in vertical equilibrium. This is because only for such flow model, the mass accretion rate 
non linearly depends on the radial sound speed, hence on the radial pressure force, which acts along the direction of the 
gravitational force, becomes more significant at any radial distance, and hence matter falls relatively less quickly, and the 
dynamical velocity can overtake the sound speed only at the close proximity of the horizon where the gravitational 
pull becomes considerably large. For conical flow, dependence of mass accretion rate on the radial distance is one power of $r$ 
more compared to $\dot{M}$ for constant height flow. The rate at which the advective velocity  will preceed the sound speed 
for a particular $r$ will thus be more for constant height flow compared to the conical flow. Similar situation is obtained 
for isothermal flow as well. There, however, the sound speed is position independent. Hence the proximity to the horizon of the sonic 
transition will solely depend on the rate at which the advective velocity increases with the radial distance. The model 
dependence of the proximity of the shock location toward the horizon follow opposite trend as is observed for the sonic points. 
Why such phenomena is observed is difficult to explain analytically sonce the shock location is determined by a set of 
non-trivially complex non linear algebro-differential equations, which are non-exactly solvable. The model dependence 
of shock related quantities can thus be demonstrated, as has been done in this work, but the exact reasons behind such 
dependencies can never be expalined analytically by any means. One can, however, make successful predictions about how several important 
features of the shocked accretion flow depends on the matter geometry. Post shock flows are more common for relatively 
slowly rotating accretion at low energy for vertical equilibrium model, whereas for the constant height model, sufficiently 
hot flow with almost Keplerian angular momentum can form shock (see e.g. figure 2). What range of initial boundary conditions 
(determined by the nature of the donor, the dynamics of the infall, the temperature of the ambient medium, etc.) 
are suitable to produce the post shock flow for which matter geometry? The answer to this important question can thus 
be obtained using our present work. The main importance of our work lies in this finding as we believe. If one knows 
the environment (which determine/is determined by the initial boundary conditions governing the flow properties) of an 
astrophysical source harbouring massive black hole (low angular momentum flow is better realised for flow onto supermassive 
black holes rather than black hole binaries), one can then predict for which kind of flow geometry the observed results 
may be explained using the multi-transonic shock accretion flow. Our work will also serve for calibration purpose to study 
the shocked black hole accretion in the source that if certain shock related property is known to have association 
with a particular type of flow geometry, then for any astrophysical source, if the observational signature of such shock 
related property is obtained, one can correctly predict the structure of the corresponding accretion disc 
using the results obtained in our work. \\

For isothermal accretion, the kind of standing shocks we consider are temperature preserving, and hence, unlike 
the polytropic Rankine-Hugoniot shock, they are dissipative. Inviscid flow may dissipate most of its binding energy at the shock. 
Such isothermal shocks may serve as the efficient provider of the effective radiative cooling mechanism 
in isothermal accretion discs. \\

\begin{figure}
\centering
\includegraphics[scale=0.7]{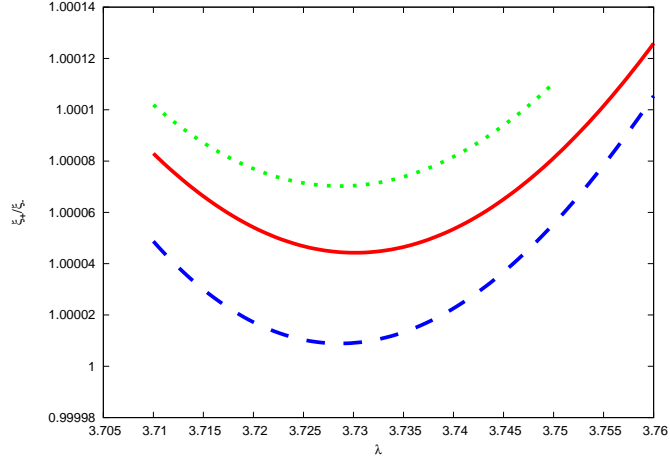}
\caption[]{$\xi_+/\xi_-$ vs. $\lambda$ for $T=10^{10}K$ 
for all three geometric configurations. Constant height disc, quasi-spherical disc and vertical equilibrium model
depicted by solid red lines, dotted green lines and dashed blue lines respectively.}
\label{fig9}
\end{figure}

A measure of the dissipation at the shock for such flow may be expressed as 
\begin{equation}
\left(\frac{\rho_+}{\rho_-}\right)^{2c_s^2}\left(\frac{1-u_-^2}{1-u_+^2}\right)
\end{equation}
where $c_s$ is the position independent isothermal sound speed which remains constant before and after a 
temperature preserving standing shock is formed. Figure (\ref{fig9}) shows the variation of such measure of dissipation 
as a function of $\lambda$ for three different flow geometries. The dissipated amount may very quickly 
be removed possibly in the form of radiation via various wavebands, most preferably in X-ray, from the disc to maintian 
the invariance of the flow temperature before and after the shock forms. Sudden X-ray burst observed from the 
galactic centre black holes may thus be explained using the formation of the dissipative shock. Isothermal 
shocks are, thus, 'brighter' compared to the polytropic Rankine-Hugoniot shock. It is observed that the isothermal 
quasi-spherical disc model may produce brightest such shock (see, e.g. figure (\ref{fig9})) whereas the 
vertical hydrostatic equilibrium disc model produces the faintest one.

\section*{Acknowledgments}
PT would like to acknowledge the kind hospitality provided by HRI, Allahabad, India, for several 
visits through the $XII^{th}$ plan budget of Cosmology and High Energy Astrophysics grant and 
SNBNCBS, Kolkata, India, for their financial and infrastructural support. 
The authors also acknowledge insightful discussions with Archan S. Majumdar. The authors would like to thank 
Sankhasubhra Nag and Sonali Saha Nag for their kind help in formulating the initial version of a numerical 
code which has been partly used in this work.

\bibliographystyle{elsarticle-harv}
\bibliography{paper}

\end{document}